\documentclass[twocolumn]{aastex62}
\usepackage[utf8]{inputenc}
\usepackage[astrosymb]{}
\usepackage{amsmath}
\usepackage{xcolor}

\graphicspath{{./}{figures/}}

\received{}
\revised{}
\accepted{December 5, 2019}

\submitjournal{ApJ}

\shorttitle{The SIBLING Survey}
\shortauthors{Mart\'inez-Palomera et al.}

\begin{document}

\title{Introducing the Search for Intermediate-mass Black-hole In Nearby Galaxy (SIBLING) Survey}

\correspondingauthor{Jorge Mart\'inez}
\email{jorgemarpa@berkeley.edu}

\author[0000-0002-0786-7307]{Jorge Mart\'inez-Palomera}
\affiliation{Department of Astronomy, University of California, Berkeley, CA, USA}
\affiliation{Department of Astronomy, University of Chile, Chile}
\affiliation{Center for Mathematical Modeling, University of Chile, Chile}
\affiliation{Millennium Institute of Astrophysics, Santiago, Chile}

\author{Paulina Lira}
\affiliation{Department of Astronomy, University of Chile, Chile}

\author{India Bhalla-Ladd}
\affiliation{Department of Physics, Yale University, CT, USA}

\author{Francisco Förster}
\affiliation{Center for Mathematical Modeling, University of Chile, Chile}
\affiliation{Millennium Institute of Astrophysics, Santiago, Chile}

\author{Richard M. Plotkin}
\affiliation{International Centre for Radio Astronomy Research, Curtin University, Perth, WA, 6845, Australia}
\affiliation{Department of Physics, University of Nevada, Reno, Nevada 89557, USA}

\begin{abstract}

Intermediate mass black holes (IMBHs) have masses between the $10^2\!-\!10^6$ M$_\odot$ and are key to our understanding of the formation of massive black holes. The known population of IMBH remains small, with a few hundred candidates and only a handful of them confirmed as bona-fide IMBHs. Until now, the most widely used selection method is based on spectral analysis. Here we present a methodology to select IMBH candidates via optical variability analysis of the nuclear region of local galaxies ($z \leqslant 0.35$). Active IMBH accreting at low rates show small amplitude variability with time scales of hours, as it is seen in one of the known IMBH NGC4395. We found a sample of $\sim \!500$ galaxies evidencing fast and small amplitude variation in their weekly based light curves. We estimate an average occupancy fraction of 4\% and a surface density of $\sim \!3$ deg$^{-2}$, which represent an increase by a factor of $\sim\!40$ compared to previous searches. A large fraction ($78\%$) of the candidates are in spiral galaxies. We preliminary confirm the AGN nature of 22 sources via BPT diagrams using SDSS legacy spectra. Further confirmation of these candidates will require multiwavelength observations, especially in X-ray and radio bands.

\end{abstract}


\section{Introduction} \label{sec:intro}

Black holes (BH) are present in the universe within a range of masses. BHs with stellar mass ($\leqslant 100$ M$_\odot$) have been detected via observation of BH-X-ray binaries \citep{2006ARA&A..44...49R} at the low mass end. More recently, gravitational waves produced by BH binary mergers have been detected by LIGO \citep[e.g.][]{PhysRevLett.116.061102, PhysRevLett.118.221101} with $M_{BH} < 50$ M$_\odot$ per-merger and up to $80 M_\odot$ for the remnant. On the other side of the mass spectrum, the existence of supermassive black holes (SMBH, $M_{BH} \geqslant 10^6$ M$_\odot$) in the center of galaxies has been proven by many studies \citep[e.g.][]{1995Natur.373..127M,2002Natur.419..694S}.
SMBH grow via accretion of gas during their active phase, via coalescence during galaxy mergers, and via tidal disruption events \citep{2002MNRAS.335..965Y,2007ApJ...665..187L,2007ApJ...667..813Y,2009MNRAS.400.2070S}. Nevertheless their formation mechanism still remains unclear. Observations of luminous quasars at redshift $z\! \sim \!7$ hosting SMBH with masses up to $\sim 10^{10} M_\odot$ \citep{2007AJ....134.2435W,2011Natur.474..616M} impose some restrictions to the formation of this SMBH and their seeds, however, they represent exceptional systems and very possibly not the norm. In the SMBH formation scenario there are mainly three possible paths \citep{2010A&ARv..18..279V,2016PASA...33...54R,2017IJMPD..2630021M}: (i) BH seeds formed from the death of first generation stars producing BHs with masses of $\sim \!10^2$ M$_\odot$; (ii) via runaway collisions/mergers of stars in dense stellar cluster producing BH seeds of $10^2 \!- \!10^4$ M$_\odot$; or (iii) from the direct collapse of large and dense cold gas in the early Universe forming BHs with masses between $10^5 \!-\! 10^6$ M$_\odot$. These processes create a population of IMBHs with masses between $10^2 \!-\! 10^6$ M$_\odot$, which are still largely undetected. Finding and studying the current local population of IMBHs, that did not evolve into SMBHs, will help to understand which seed formation/s scenario/s is/are correct.

Using SDSS spectra, several authors have found hundreds of IMBH candidates by looking for broad components to the Balmer lines observed in galactic nuclei  \citep{2004ApJ...610..722G,2007ApJ...670...92G,2012ApJ...755..167D,2013ApJ...775..116R,2018ApJS..235...40L,2018ApJ...863....1C}. The known scaling relations between the size of the region that produces these broad lines and the optical continuum luminosity of the active nucleus enable them to determine the Viral BH mass from these Active Galactic Nuclei (AGN) as $\sim \!R v^2/G$, where $R$ is the size of the emitting region, $v$ is its orbital velocity (obtained from the width of the broad lines) and $G$ is the gravitational constant. These `single-epoch' mass estimations are mainly found at the high end of the mass range of IMBHs and also at high Eddington ratio.

One of the most remarkable known IMBHs is the one present in NGC4395. This BH was repeatedly observed by \citet{2012ApJ...756...73E} in the optical wavelengths during nine nights producing well sampled light curves with 15 min cadence. Their Figure 2 shows the typical variability of this IMBH with peak-to-peak amplitudes below $0.2$ magnitudes, clear structure within the photometric uncertainties, and characteristic time scales of a couple of hours. If optical variability from IMBHs is due to reprocessing of a rapidly varying high-energy signal originating close to the BH \citep[e.g. ][]{2008ApJ...689..762D,2012ApJ...751...39K}, the shortest optical variability time scales are expected to depend on the size of the accretion disk. For example, assuming a Shakura-Sunyaev \citep{1973A&A....24..337S} accretion disk and relating its temperature to a characteristic wavelength using the Wien displacement law, we can estimate the size of the accretion disk in NGC4395 as $R \sim 5 \times 10^2$ light-seconds for a IMBH of $\sim \!10^5$ M$_\odot$. This shows that optical variability faster that $2 \times R \sim 0.3$ hours should not be observed, which matches well the observations. Following this, \cite{2012ApJ...751...39K} selected a sample of 15 IMBH candidates in the $(1.1-6.6) \times 10^6 M_\odot$ mass range by variability and spectral analysis in the X-rays.

It its well known that AGN are variable sources \citep[e.g.,][]{1963ApJ...138...30M,2001sac..conf....3P}, and this behaviour has been extensively used to select AGN candidates. Recently \cite{2018ApJ...868..152B} used SDSS Stripe 82 data to identify 135 AGNs in low-mass galaxies via long-term optical variability. Because of the low-mass of the hosts, it is expected that these sources harbour IMBHs. However, as shown by NGC4395, IMBHs should show fast, intranight variability and hence, a high cadence variability survey is necessary to efficiently find these sources. Combining their characteristic variability and expected time-scales for accreting IMBH, here we present a search for IMBH via optical variability using relatively high-cadence observations: the Search for Intermediate-mass BLack-holes In Nearby Galaxies (SIBLING) survey.
Taking advantage of the High Cadence Transient Survey \citep[HiTS, ][]{2016ApJ...832..155F}, a remarkable survey that combines wide, deep and fast cadence observations, enable us to perform a non-targeted search for IMBH selected by short-term variability on their light curves.

This article is structured as follows: in Section \ref{sec:data} we present the data used for this study, sample selection and photometric procedures; in Section \ref{sec:lc} we present the variability selection criteria and long-term analysis; Section \ref{sec:varCand} presents properties of the variability selected catalog; finally we give our conclusions and final thoughts in Section \ref{sec:concl}.

\section{Data Processing} \label{sec:data}

\subsection{HiTS data} \label{subsec:HiTSdata}

The HiTS is an imaging survey, using the Dark Energy Camera (DECam) at the 4m Blanco telescope on Cerro Tololo Interamerican Observatory (CTIO). It consist of three observational campaigns during 2013, 2014 and 2015. One week of relatively high cadence, large sky coverage and high limiting magnitude observations were conducted during each year. HiTS was designed to study the early phases of supernova events \cite[e.g.][]{2016ApJ...832..155F,2018NatAs.tmp..122F}. Its unique specifications enabled also to search for distant RR-Lyrae \citep{2017ApJ...845L..10M,2018ApJ...855...43M}, asteroids \citep{2018AJ....155..135P}, and Machine Learning classification \citep{2017ApJ...836...97C,2018AJ....156..186M}.

In this work, we have used data from the 2014 and 2015 campaigns. These consist of imaging data near the equatorial plane \citep[see Figure 4 in][]{2016ApJ...832..155F} mainly in the \textit{g} band. Observations during 2014 covered 120 deg$^2$ in 40 fields during five consecutive nights, with exposure times of 160 seconds. Each field was observed four times per night, giving an observing cadence of 2 hours. During 2015, observations covered 150 deg$^2$ in 50 fields during six nights, with exposure times of 86 seconds. Each field was observed five times per night, giving a cadence of 1.6 hours. There is an overlap of 42 deg$^2$ (14 DECam fields) between both years. The typical limiting magnitude of HiTS is $\sim 24.3$ in the \textit{g} band \citep{2018AJ....156..186M}.

Images were pre-processed by the DECam Community Pipeline \citep{2014ASPC..485..379V} and delivered by the NOAO archive.

\subsection{Sample selection}\label{subsec:sample_selection}

In order to select a sample of nearby galaxies available in our dataset, we performed a cross-match between the catalog of detected sources in the HiTS survey \citep{2018AJ....156..186M} and the Sloan Digital Sky Survey Data Release 12 catalog \citep[SDSS DR12, ][]{SDSS-DR12}. The SDSS catalog contains a total of $2\,401\,952$ spectroscopically confirmed galaxies and $477\,161$ quasars, distributed in an area of $9\,376$ deg$^2$. We filtered all the sources with spectroscopic class (\textit{spCl}) equal to GALAXY. This gave us a sample of $6\,703$ and $7\,592$ objects for 2014 and 2015, respectively. The SDSS only covers down to $-4$ deg in declination, therefore the overlap with HiTS 2014 and 2015 is not complete, with 26 HiTS fields available for 2014 and 30 fields available in 2015. Thus, the efective search area is $\sim \! 168$ deg$^{-2}$. In what follows we refer to the galaxy sample from the two years separately as the galaxy sample. The SDSS catalog also contains two other classifications for the GALAXY class: the first one is the source type, \textit{spType}\footnote{\url{http://cdsarc.u-strasbg.fr/ftp/cats/V/147/type9.htx}\label{fn:sptype}}; and the second one is the spectroscopic subclass (e.g. BROADLINE, STARFORMING, STARBURST, and AGN). The former is dominated by three groups of sources due to the SDSS target selection: bright galaxies from the Main Galaxy Sample \citep{2002AJ....124.1810S}, which account for $\sim1/3$ of the sample; Luminous Red Galaxies \citep[LRG, ][]{2001AJ....122.2267E}, which accounts for $\sim2/3$ of the sample; and a small fraction ($\sim 2\%$) of low redshift quasars. We cleaned the galaxy sample by removing all sources without spectroscopic redshifts and kept only the magnitude range covered by HiTS ($15 < m_g < 25$). This reduces our galaxy sample by less than 0.5\%.

\subsection{Photometry and Light Curve Construction} \label{subsec:reduction}

To determine the nuclear light curves it is necessary to measure the flux independently of the seeing conditions of each observation, as seeing variations change in a different manner the amount of galaxy light (a resolved sources) and light from an active nucleus (an unresolved source) contained in a given aperture. To achieve this two approaches are usually taken: 1) through convolution downgrade all images to a common seeing and use a fixed aperture to measure the nuclear flux; 2) build a deep master frame using images obtained under good seeing conditions, convolve it to match the seeing of each individual observation, scale it in flux and match its astrometry, and finally subtract it from each image and obtain forced photometry at the position of the galactic nuclei. This last method takes advantage of the intrinsic seeing condition of each observation, and therefore it delivers a deeper resulting photometry. The first method, on the other hand, forces all observations to be downgraded in seeing but it is more straightforward and hence more robust. For this work we choose to use the first described method.

For each year we selected all the observations with measured seeing $ \leqslant 1.8$ arcsec and observed airmass $\leqslant 2$ to remove images with poor observation conditions. This restriction removes $26 \%$ of the images from the 2014 campaign, and the light curves of sources observed during this year have at most twenty observations. On the other hand, due to bad observing conditions we only kept data from the first three nights of observations in 2015. Hence, light curves calculated for sources observed during this year have at most fifteen observations. Seeing filtering was not needed in this case due to the excellent observing conditions of the remaining three nights of the 2015 campaign.
Afterwards, we extracted $51\times51$ pixel stamps for each galaxy selected above. From the characterization of point-like sources observed in the same field as our targeted galaxies, we selected the stamp from the image with the largest seeing and used it as reference epoch. The remaining stamps were converted to match its seeing. The convolution kernel was calculated following the method described in Section 4.1 of \citet{2016ApJ...832..155F}. Appendix Section \ref{apx: img_lc_ex} show several examples of the original and convolved images, as well as the convolution kernel for a selection of galaxies.

After seeing matching we performed aperture photometry at the centroid position of each galaxy nucleus. Aperture radii were set as $0.5, 0.75, 1.0, 1.25$ and $1.5$ times the seeing of the reference image. These gave us typical apertures between $0.6$ and $2.2$ arcsec. Photometric calibrations were calculated following the same procedures described in \cite{2018AJ....156..186M}. In particular, we used the zero points calculated against PanSTARRS catalogs for every Field/CCD/epoch combination, and photometric uncertainties were corrected by an empirically determined pixel correlation.

The above procedures provided light curves for each galaxy in the sample. We constructed light curves using all the apertures  described before.
We tested our photometric procedure constructing light curves for a set of non-variable stars taken from \cite{2018AJ....156..186M}. The typical standard deviation of these light curves was $0.02$ mags for the \textit{g} band, which is below the typical photometric uncertainties. A light curve example of a non-variable star is shown in Figure \ref{fig:lc_examples} panel (a).

\begin{figure}[ht!]
\gridline{\fig{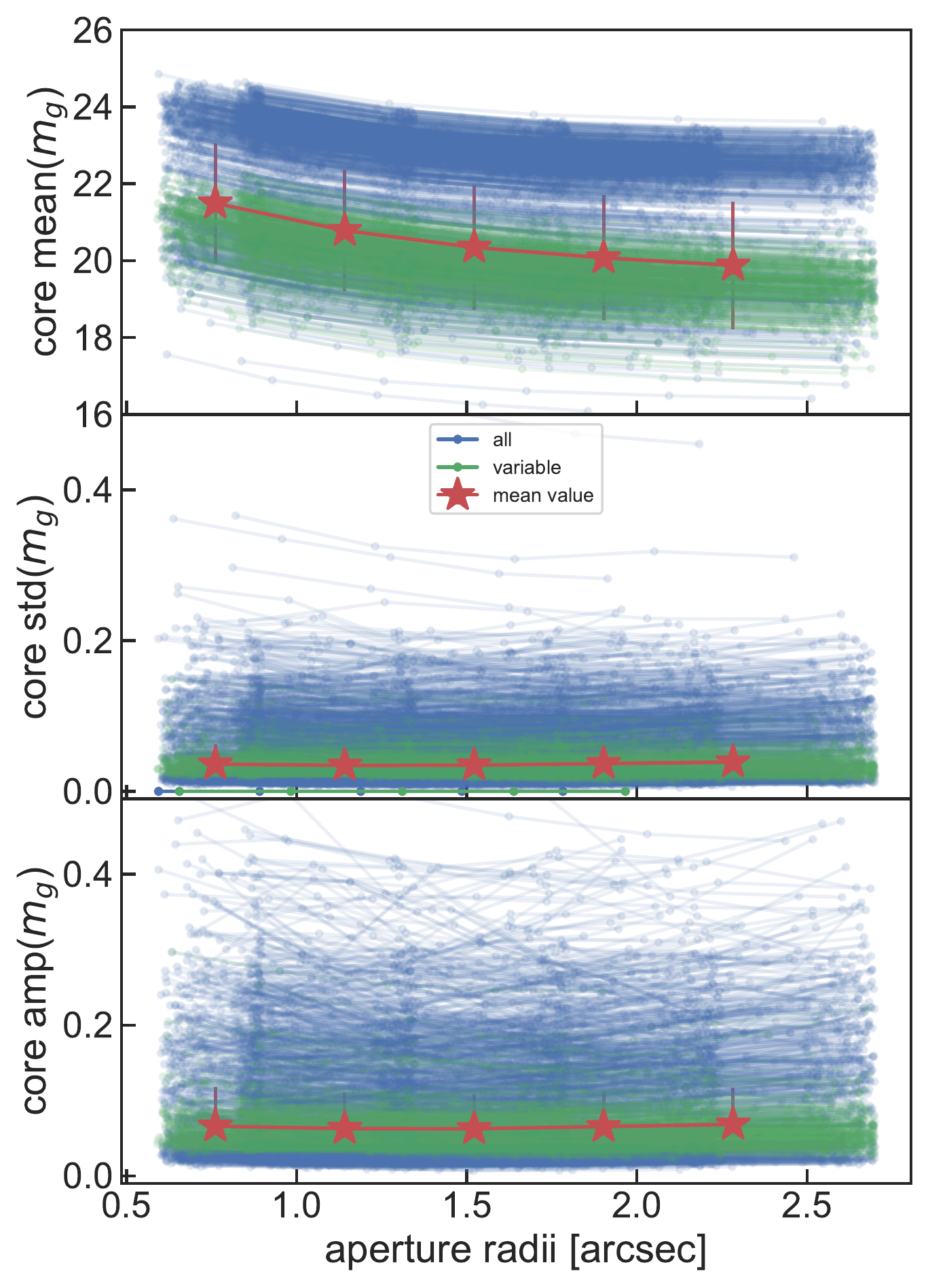}{0.45\textwidth}{}}
\caption{Mean magnitude (top), standard deviation (middle), and amplitude (bottom) of a sample of light curves as a function of time (blue lines). Blue lines show non-variable galaxies, while in green lines are galaxies selected as variables, see Section \ref{sec:lc}. Red markers and error bars represent the median and the median-absolute-deviation of the sample, respectively, per bin of aperture.}
\label{fig:feat_ap}
\end{figure}

We analyzed two variability indicators, amplitude and standard deviation, as a function of the photometric aperture used to construct the light curves. Although there is clear correlation between the measured flux and the size of the aperture, we found (on average) no significant correlation between the variability indicators and the aperture size (see Figure \ref{fig:feat_ap}). A small number of sources show dependency with the used aperture, these correspond to sources with bad photometry, contamination from close sources, and/or kernel convolution problems. We also inspected the measured flux after convolution as a function of the seeing before convolution for a sample of non-variable sources. We found that there is no residual correlation between these two quantities for all different photometric apertures used. Therefore, there is no preference for aperture choice regarding variability. We decided to use an aperture size equivalent to the seeing at the reference image for each light curve to avoid flux losses. This also helps to securely enclose the core region of the wide variety of morphological classes present in the sample. Figure \ref{fig:lc_examples} shows images of two galaxies and the used aperture size.

\section{Variability Analysis} \label{sec:lc}
\subsection{Selection Criteria}

We calculate a set of variability features for each light curve: \textit{Amplitude}, \textit{ExcessVariance} ($\sigma^2_{rms}$), \textit{MedianAbsDev}, and \textit{MedianErr}. We have used the excess variance definition adopted by \cite{2017ApJ...849..110S}:

\begin{equation} \label{eq:excessVar}
\sigma^{2}_{rms} = \frac{1}{N_{obs}\bar{x}^2} \sum^{N_{obs}}_{i=1}{[(x_i - \bar{x})^2 - \sigma^2_{err,i}]}
\end{equation}

and its uncertainty due to Poisson noise:

\begin{equation} \label{eq:excessVar_err}
err(\sigma^2_{rms}) = \frac{S_D}{\bar{x}^2 N^{1/2}_{obs}}
\end{equation}

\begin{equation} \label{eq:Sd}
S_D = \frac{1}{N_{obs}} \sum^{N_{obs}}_{i=1}{\lbrace[(x_i - \bar{x})^2 - \sigma^2_{err,i}] - \sigma^2_{rms}\bar{x}^2\rbrace^2}
\end{equation}

where $N_{obs}$ are the number of data points in the light curve, $\bar{x}$ is the mean magnitude, and $x_i$ and $\sigma_{err,i}$ are the magnitude and its error at each epoch, respectively. If the value $(\sigma^2_{rms} - err(\sigma^2_{rms})) > 0$, then the source has detected variability.

First, we select intrinsically variable sources by filtering all the light curves with $ExcessVariance > 0.001$, a value similar to the standard deviation of all positive values for this feature. Although a positive value is already a sign of intrinsic variability, we decided to narrow down the selection in order to secure the level of variability. This reduces our galaxy sample from $6\,703$ and $7\,592$ to $1\,445$ and $1\,360$ ($\sim\! 21\%$ and $\sim\! 18\%$) for 2014 and 2015, respectively.
Since we also expect small amplitude variability, we filtered by $Amplitude < 0.1$. This threshold is based on the characteristic amplitude found for NGC4395, which is of the order of 0.2 peak-to-peak magnitudes in the \textit{g} band. Through visual inspection of light curves, we found that this cut also helps to remove sources with possible variability due to bad photometry or kernel convolution problems which usually introduce sharp, high-amplitude variations in the light curves. This reduces our samples to $1\,346$ and $1\,234$, which represents $\sim\! 20\%$ and $\sim\! 16\%$ of the galaxy sample. 
In order to secure the significance of the detected variability we filtered by the median photometric uncertainty of the light curve to $MedianErr < 0.05$. This limit is slightly larger than the median value of its distribution. This reduces our sample to $497$ and $412$ sources for each year.
Finally, to narrow down the selection further we filtered by the median absolute deviation value of the magnitude distribution to $MedianAbsDev > 0.015$. Again, the cut was chosen as the median of its distribution. This leads us to a sample of $302$ and $229$ sources for the 2014 and 2015 data, respectively, which are selected as variable galaxy nuclei with rapid, significant variability on the scale of hours.

\begin{figure*}[ht!]
\gridline{\fig{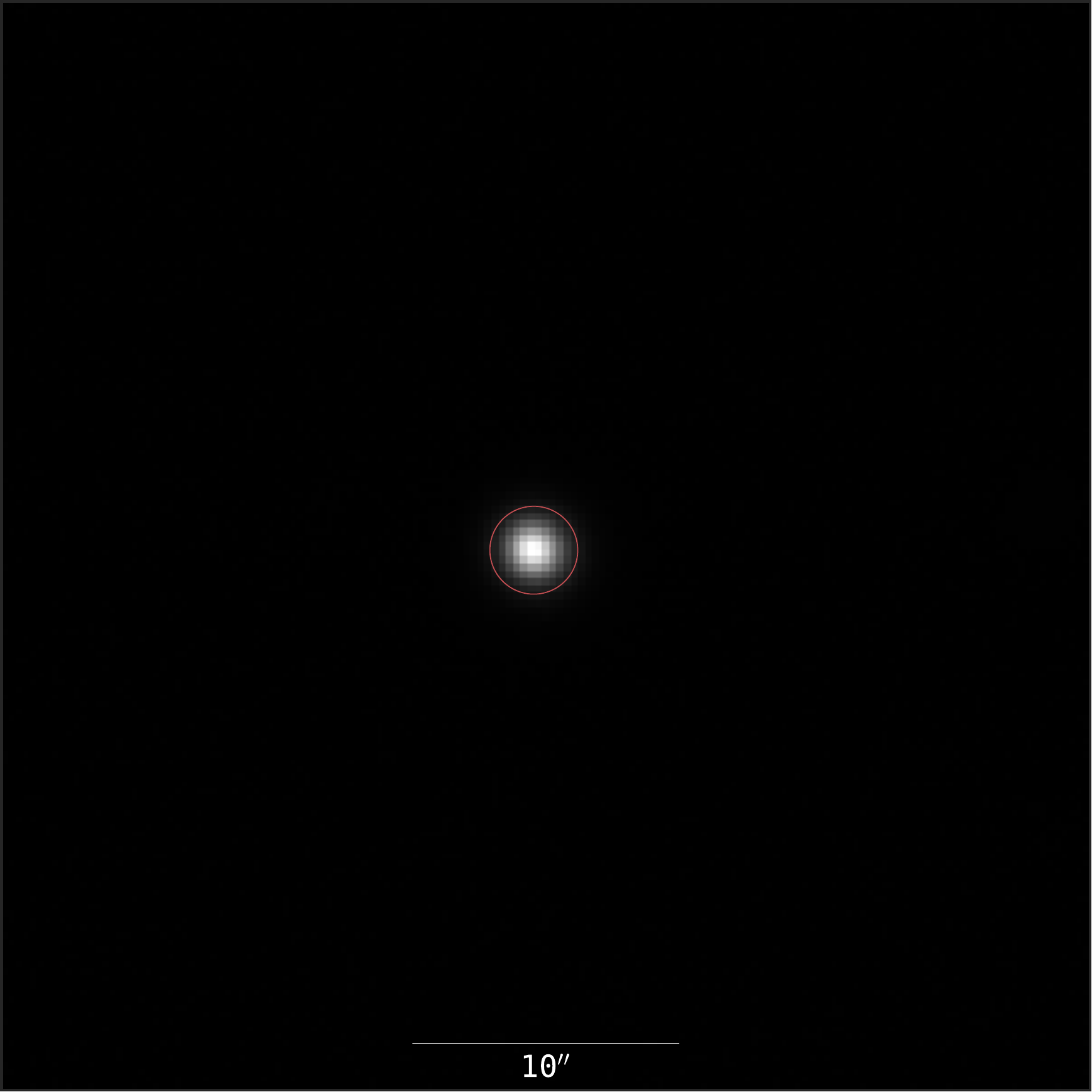}{0.23\textwidth}{}
          \fig{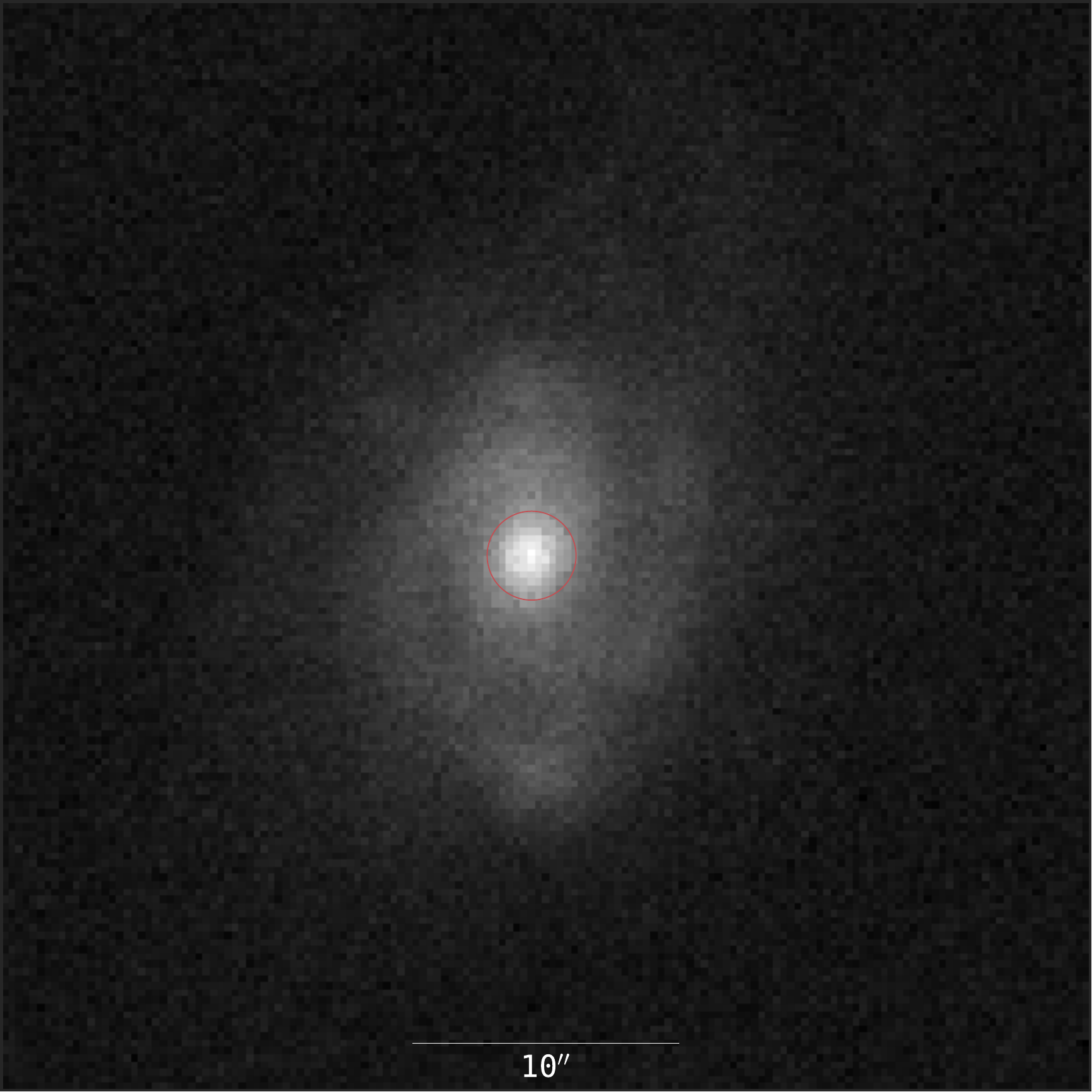}{0.23\textwidth}{}
          \fig{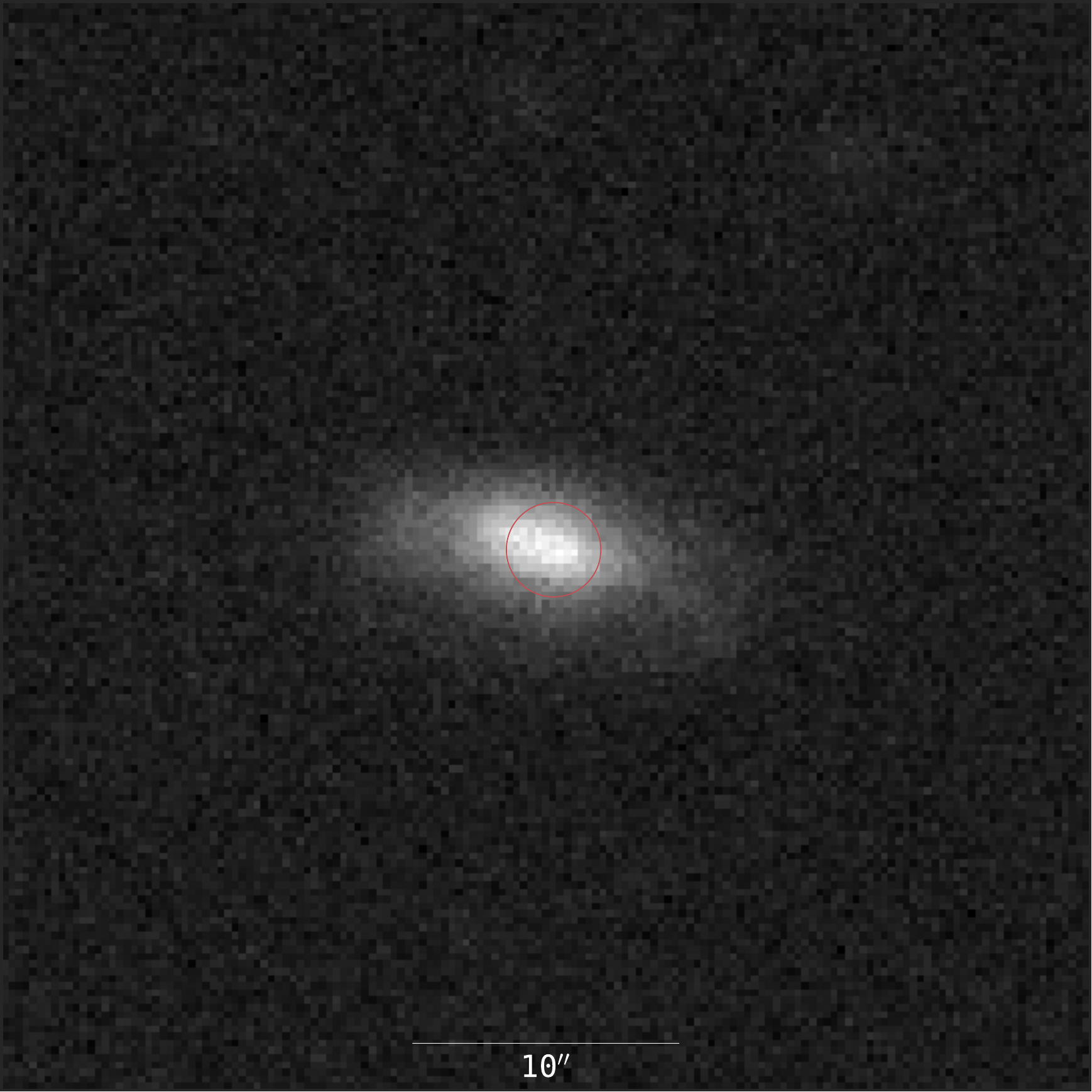}{0.23\textwidth}{}}
\gridline{\fig{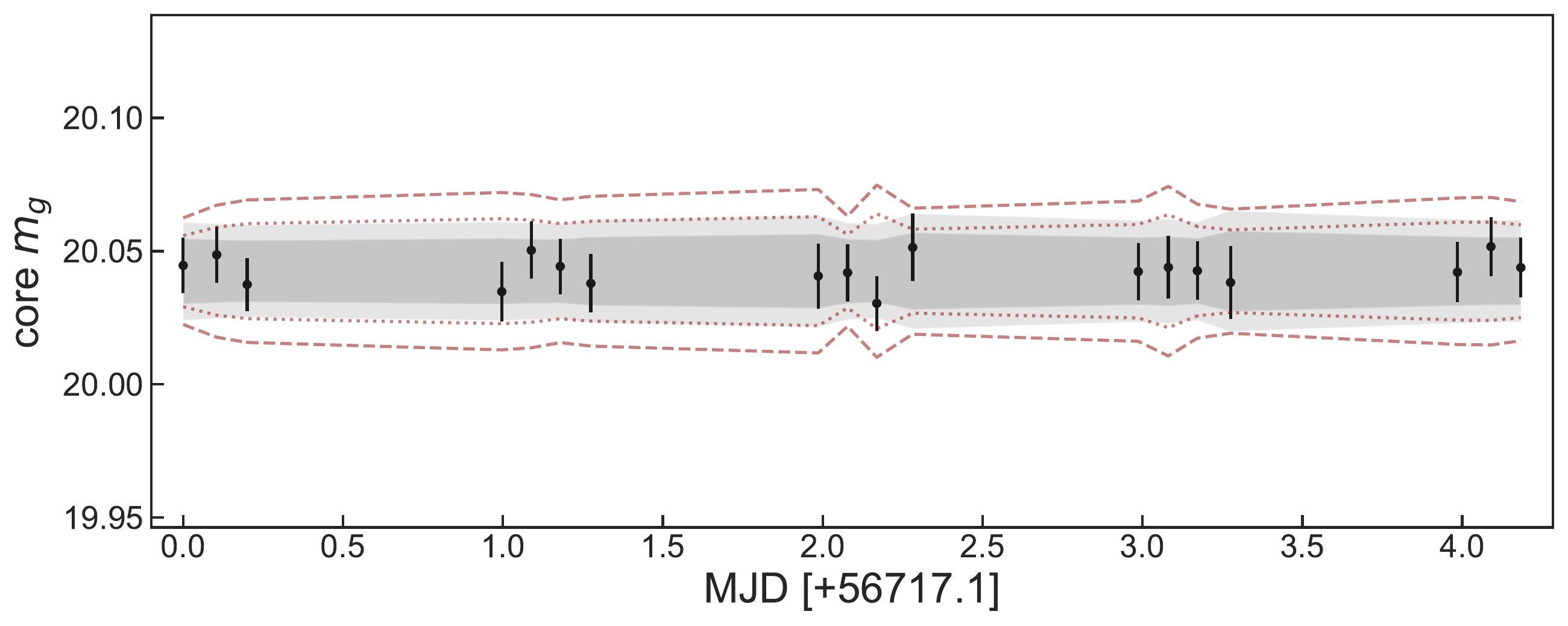}{0.33\textwidth}{(a)}
          \fig{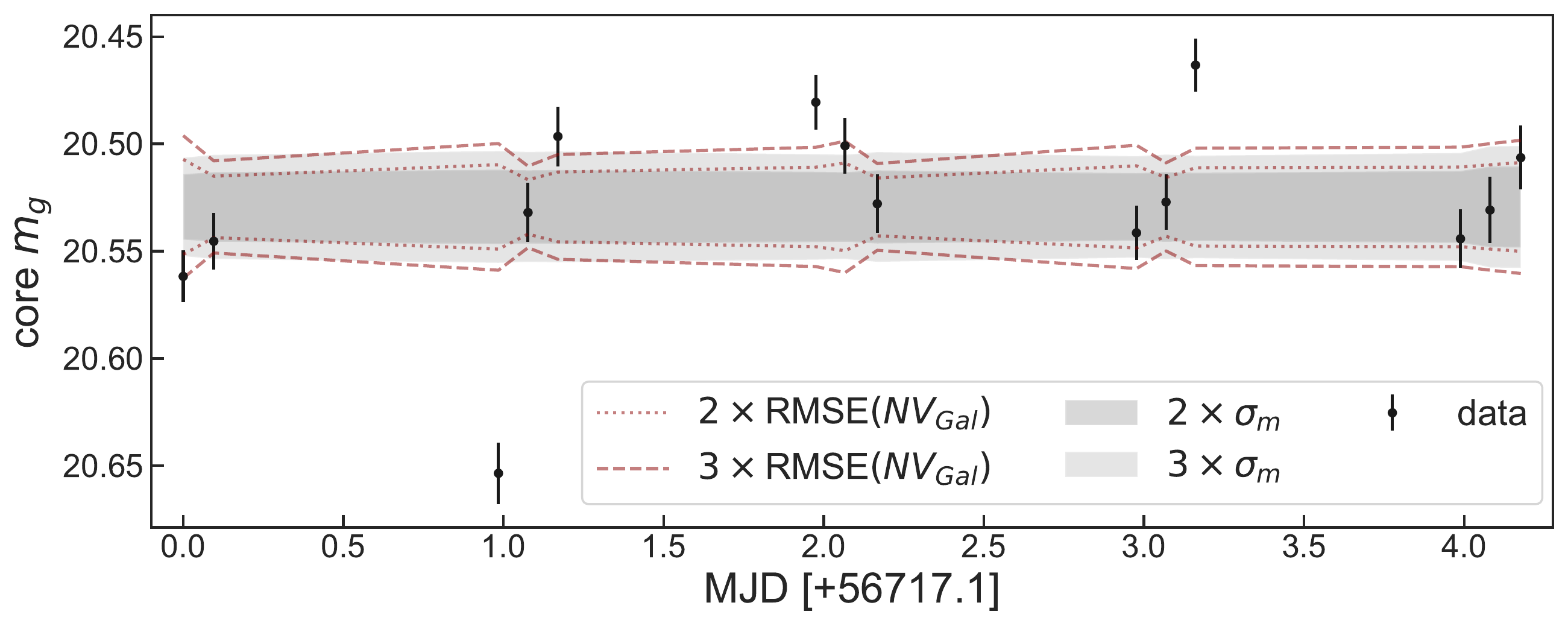}{0.33\textwidth}{(b)}
          \fig{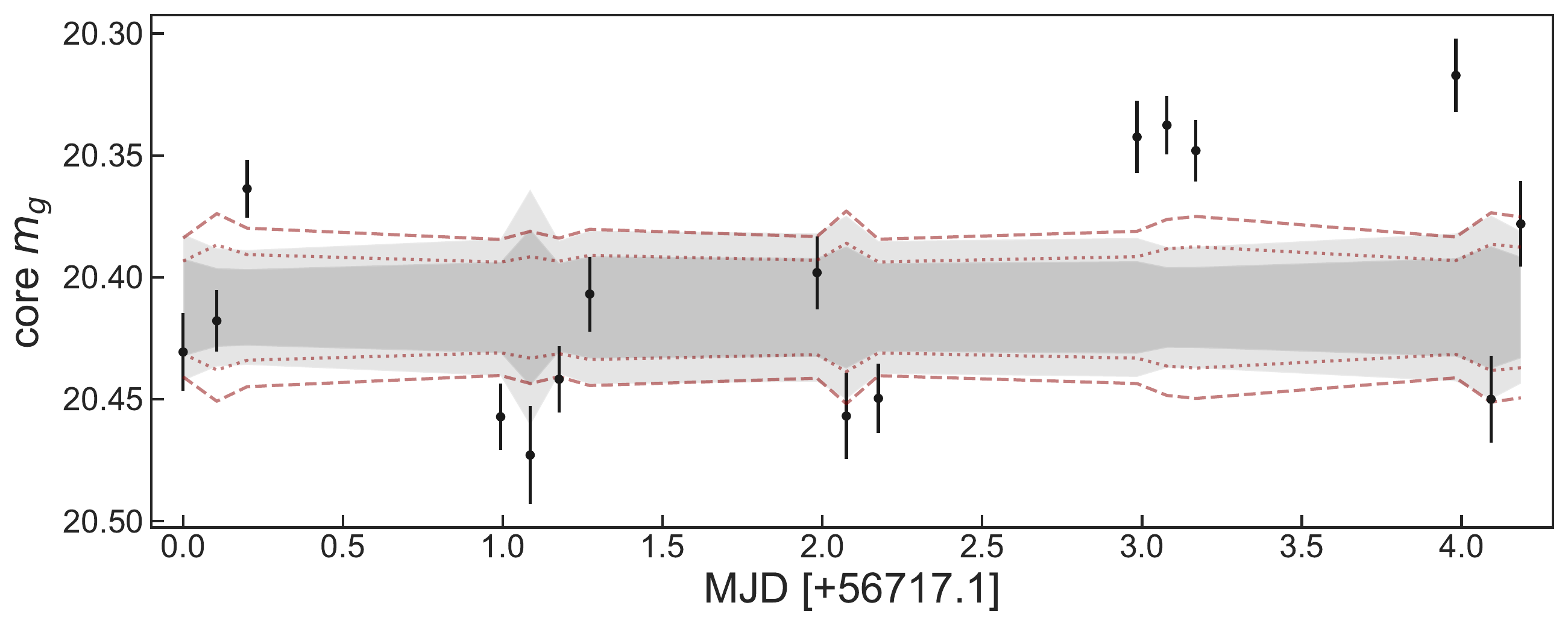}{0.33\textwidth}{(c)}}
\caption{Examples of DECam images (top row) and light curves (bottom row) produced by our analysis. Panel (a) is a non-variable star in the field, while the panels (b) and (c) are galaxies which meet our variability selection criteria. Images are 40 arcseconds across and have superimposed the aperture used for the photometry, which is equal to the seeing of the reference image. The bottom row shows the nuclear \textit{g} band light curves. The black points are the observed magnitudes with corresponding uncertainties. Red dashed lines represent the 2 and 3 sigma intervals for non-variable galaxies in the field, while the gray shaded regions shows the 2 and 3 times the photometric uncertainties around the median magnitude.
\label{fig:lc_examples}}
\end{figure*}

Figure \ref{fig:lc_examples} shows two examples of galaxies with variable nuclei from our selection as well as an example of a non-variable star in the field. More light curve examples of selected variable galaxies are shown in the Appendix Section \ref{apx: img_lc_ex}. We inspect all light curves in our variable selected sample and compared their variability against non-variable galaxies ($ExcessVariance < 0$, shown in Figure \ref{fig:lc_examples} with red dashed lines for 2 and 3 sigma intervals) to ensure the significance of the detection.

\subsection{Completness}

Thanks to the overlap between the fields observed during the 2014 and 2015 HiTS observational campaigns, we have the chance to study the fraction of selected sources after a one year lapse. We correlate the galaxies selected as variable and non variable in both years. There are 1\,986 SDSS galaxies available in the 14 overlapping fields, from which 110 and 78 galaxies were selected as variables in 2014 and 2015, respectively. 29 sources were selected as variables in both years. 81 sources were selected as variables in 2014 but not in 2015, and 49 sources were selected as variables in 2015 but not in 2014.

\begin{figure}[h!]
\includegraphics[width=.47\textwidth]{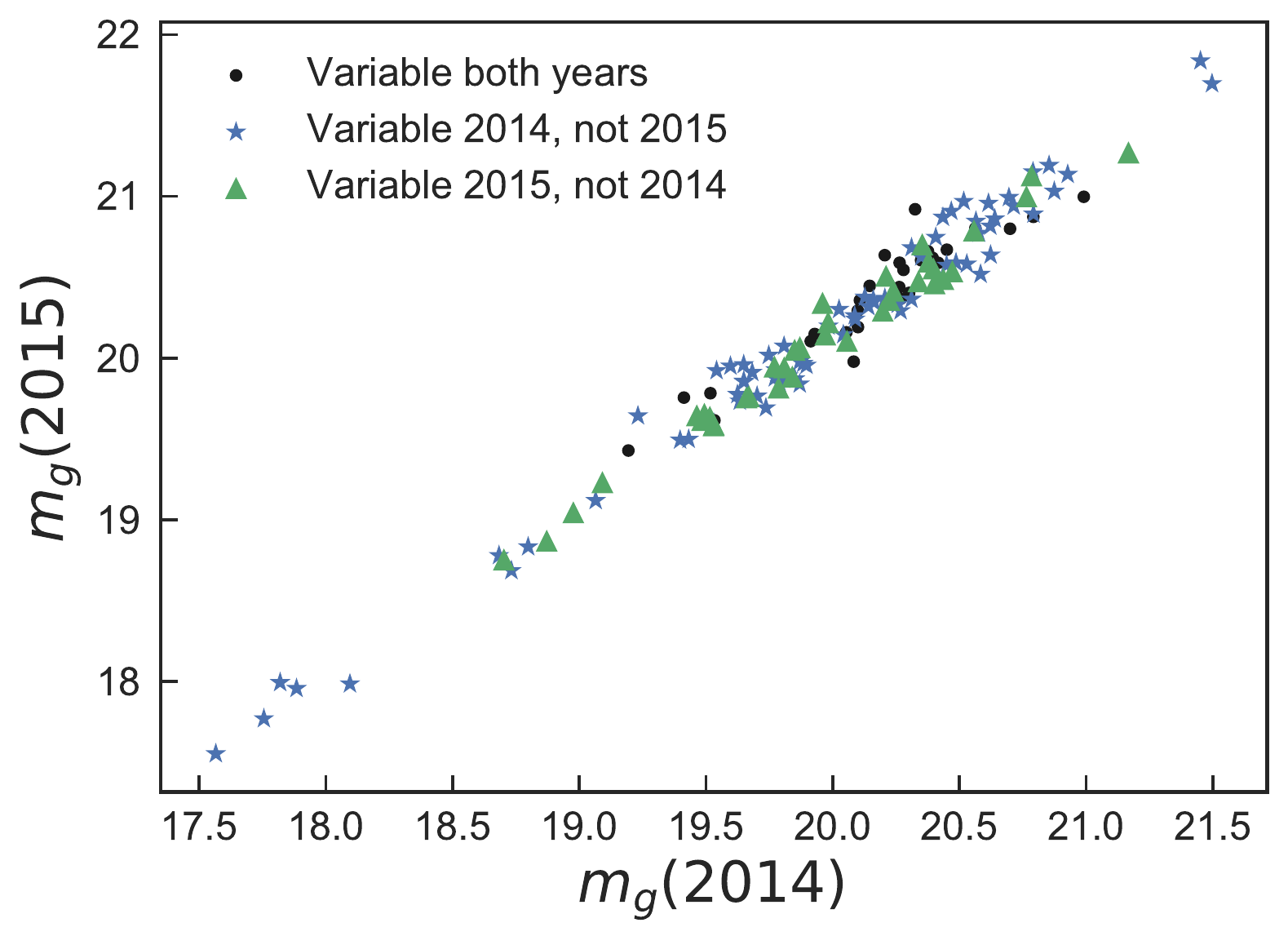}
\includegraphics[width=.48\textwidth]{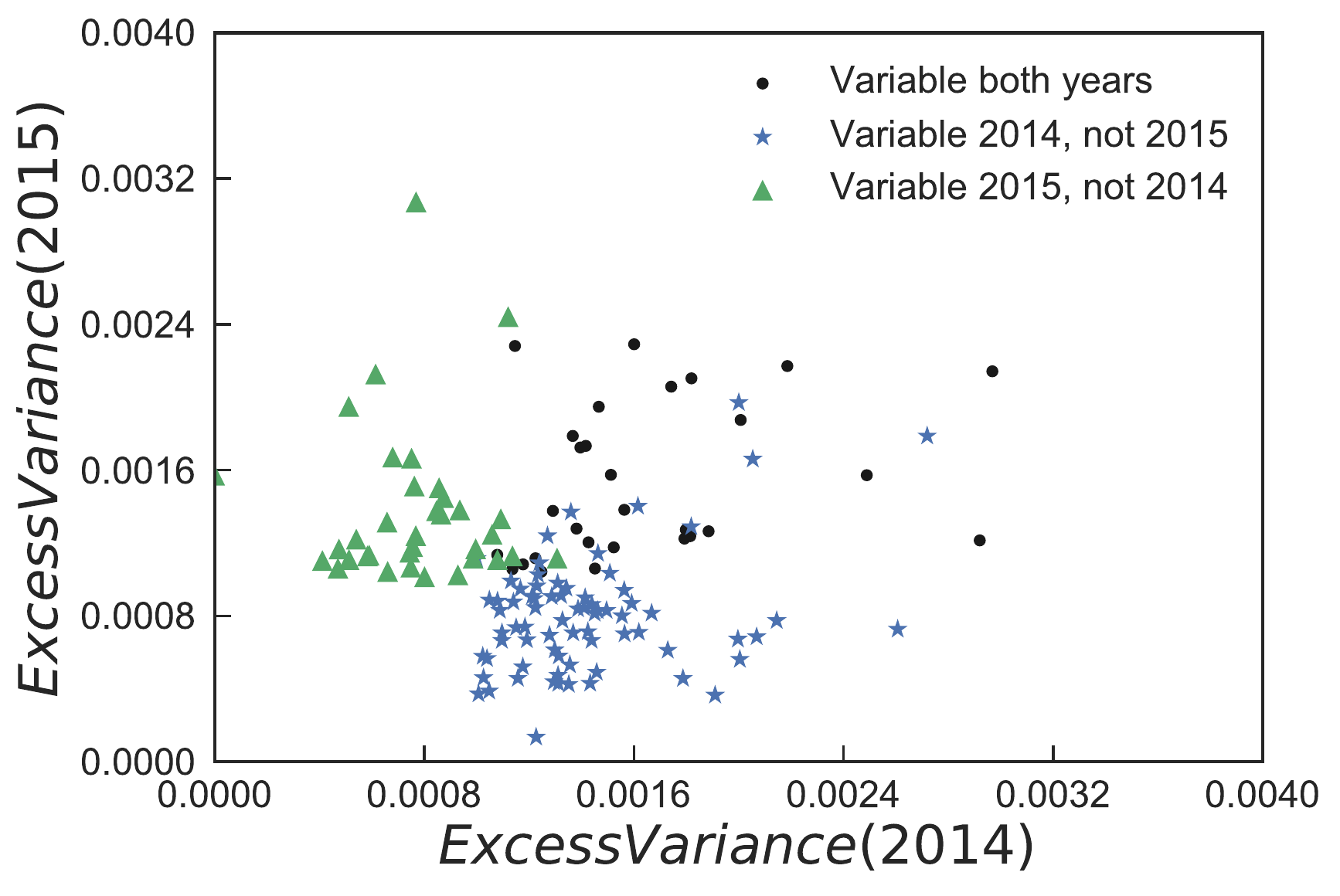}
\caption{Mean core magnitude (top panel) and $ExcessVariance$ (lower panel) from 2014 and 2015 data. Black dots are sources selected as variable in both years. Blue stars are galaxies selected as variables in 2014 only. Green triangles are variables selected in 2015 only.
\label{fig:14v15_var}}
\end{figure}

Figure~\ref{fig:14v15_var} shows feature values of mean core magnitude $m_g$ (panel a), and $ExcessVariance$ (panel b), calculated for galaxies with both 2014 and 2015 data. In black dots we show those sources selected as variable both years (30), with blue stars those variables on 2014 but not in 2015 (81), and with green triangles those for the opposite case (49). It can be seen that the mean magnitude of the galaxy nuclei during 2014 and 2015 are well correlated for the three samples. Therefore, detections in one year and non--detections in the other, are not due to severe changes in source brightness. Instead, the amplitude of the variability changes significantly as can be seen in panel (b), possible due to actual changes in the variability properties of these sources, which is still not known for the IMBH regime.
Moreover, all galaxies which show variability during 2015 only have an excess variance $\leqslant 0.001$ during 2014 and therefore were excluded after applying the variability filters.
A few sources which are only variable in 2014, on the other hand, have excess variance $>0.001$ in 2015. These sources are not picked up as variables in 2015 data because they not surpass the $MedianAbsDev$ filter. Sources selected as variables during both years are located in the middle of the two samples in the $ExcessVariance$ plot.

During 2014 we select as variables $\sim5\%$ of the total galaxies available (1\,986) in the 14 fields shared between both years. This number drops to $\sim3\%$ in 2015. If we assume that all the selected variable galaxies during both years are true variables and missing them is only related to the quality of our data, namely the length of the light curve or the image quality, we can estimate the fraction of variables recovered each year as $\sim 70\%$ during 2014, and $\sim 50\%$ during 2015. Thus, we conclude that during 2014 we were more efficient recovering variable galaxy nuclei. Then, having slightly longer light curves (remembering that for 2014 we used the 5 days of observations available giving us 20 data points, compared to 15 epochs from the three first days in 2015) and deeper observations (160 second of exposure time during 2014 versus 86 seconds during 2015) improve the number statistics, and therefore, the variability selection. This also shows that we are, at least $\sim 50-70\%$ complete in our candidate selection to the typical depth of our survey.

\section{Results} \label{sec:varCand}

Combining both years gives a total of 502 unique galaxies at $z < 0.34$. This translates into a fraction of $502/12300 \sim 0.04$ candidate IMBH per galaxy, with a number density of $502/168$ deg$^{-2} \sim 3$ candidate IMBH per deg$^{-2}$. Given the strict variability constraints imposed before, this should be regarded as a lower limit to the number of IMBH candidates. Besides, comparing the positive detections in those fields that where observed during both 2014 and 2015 campaigns show that about 25\%\ of the nuclei were selected in one year but not in the other due to a lower variability amplitude.

\citet{2018ApJS..235...40L} found over 500 low-mass black holes in the $(1-20)\times 10^5 M_{\odot}$ mass range by looking at broad Balmer emission lines. Their host galaxies are dominated by young stellar populations, with masses in the range of $10^{8.8} - 10^{12.4} M_{\odot}$, and with typical optical colors consistent with spirals. When comparing our results with those of \cite{2018ApJS..235...40L} we find that three (out of 8) of their IMBH candidates in the HiTS footprint were selected by our search. Taking into account that these are completely independent methods, it is found that recovering three sources is significant, since the probability of finding these three sources is less 0.8\%\ for a random draw of 502 objects from the complete pool of available galaxies to both surveys (which corresponds to the HiTS footprint, since \cite{2018ApJS..235...40L} cover the whole SDSS).

\subsection{Host Properties}

\begin{figure*}[ht!]
\includegraphics[width=1.\textwidth]{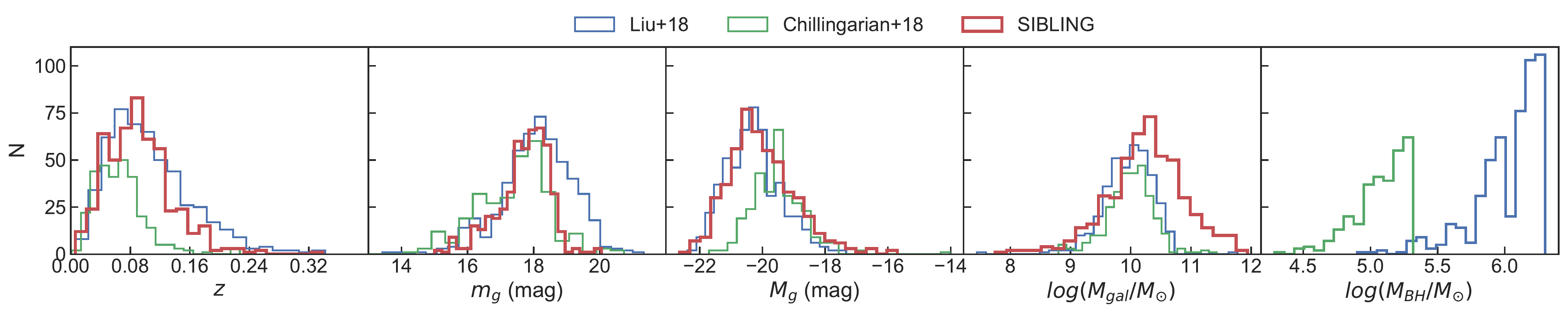}
\caption{Distribution of redshift, apparent and absolute \textit{g} magnitude, and mass of candidate IMBH host galaxies from this work (red). The value distribution from \cite{2018ApJS..235...40L} and \cite{2018ApJ...863....1C} are shown in blue and green lines for comparison. The BH mass distribution from these two works are also shown in the right panel.
\label{fig:varGal_hist_papers}}
\end{figure*}

The galaxies hosting candidate IMBHs have $-22.65 < M_g < -15.7$, $7.7 < log(M_{gal}/M_{\odot})<12$ \citep[][\footnote{\url{https://www.sdss.org/dr14/spectro/galaxy_portsmouth/}}]{2009MNRAS.394L.107M}, and are found out to redshift $0.34$. The mean host mass and absolute magnitude are $1.7 \times 10^{10} M_\odot$ and -20.1 magnitudes, respectively. No variable nuclei were associated to galaxies from the SDSS LRG sample. Figure~\ref{fig:varGal_hist_papers} presents the redshift, apparent and absolute magnitude, and galaxy mass distributions of our sample and those of \cite{2018ApJS..235...40L} (complemented with the sample selected by \cite{2012ApJ...755..167D}, leading to a total of 513 candidates) and \cite{2018ApJ...863....1C} (total of 305 candidates). Galaxy masses were also obtained from \cite{2009MNRAS.394L.107M}, having 100\% cross-match for \citeauthor{2018ApJ...863....1C} sample, and $\sim \! 70\%$ cross-match for \citeauthor{2018ApJS..235...40L} sample. The missing 30\% in the latter compromise the high-end of the luminosity distribution, which translate in the under-representation of high-mass systems, $log(M_{gal}/M_{\odot})>10.5$. This explains the mismatch between our variability selected sample and \citeauthor{2018ApJS..235...40L} sample in the galaxy mass distribution, although the redshift and luminosity distributions are in agreement. Both authors used SDSS spectra to search for IMBHs detected through the presence of weak, broad components to the Balmer lines. The distribution of BH masses derived from their samples using single-epoch calibrations is also shown.

We combine morphological information from Galaxy Zoo (GZ) catalogs and our own visual classification in order to asses galaxy types for uncertain GZ classifications. From the 502 variable galaxy nuclei previously selected 392 are spirals, 68 elliptics, 30 irregulars, 7 mergers, 2 quasars, and 3 are undetermined from SDSS and DECam images.

\begin{figure}[ht!]
\includegraphics[width=.45\textwidth]{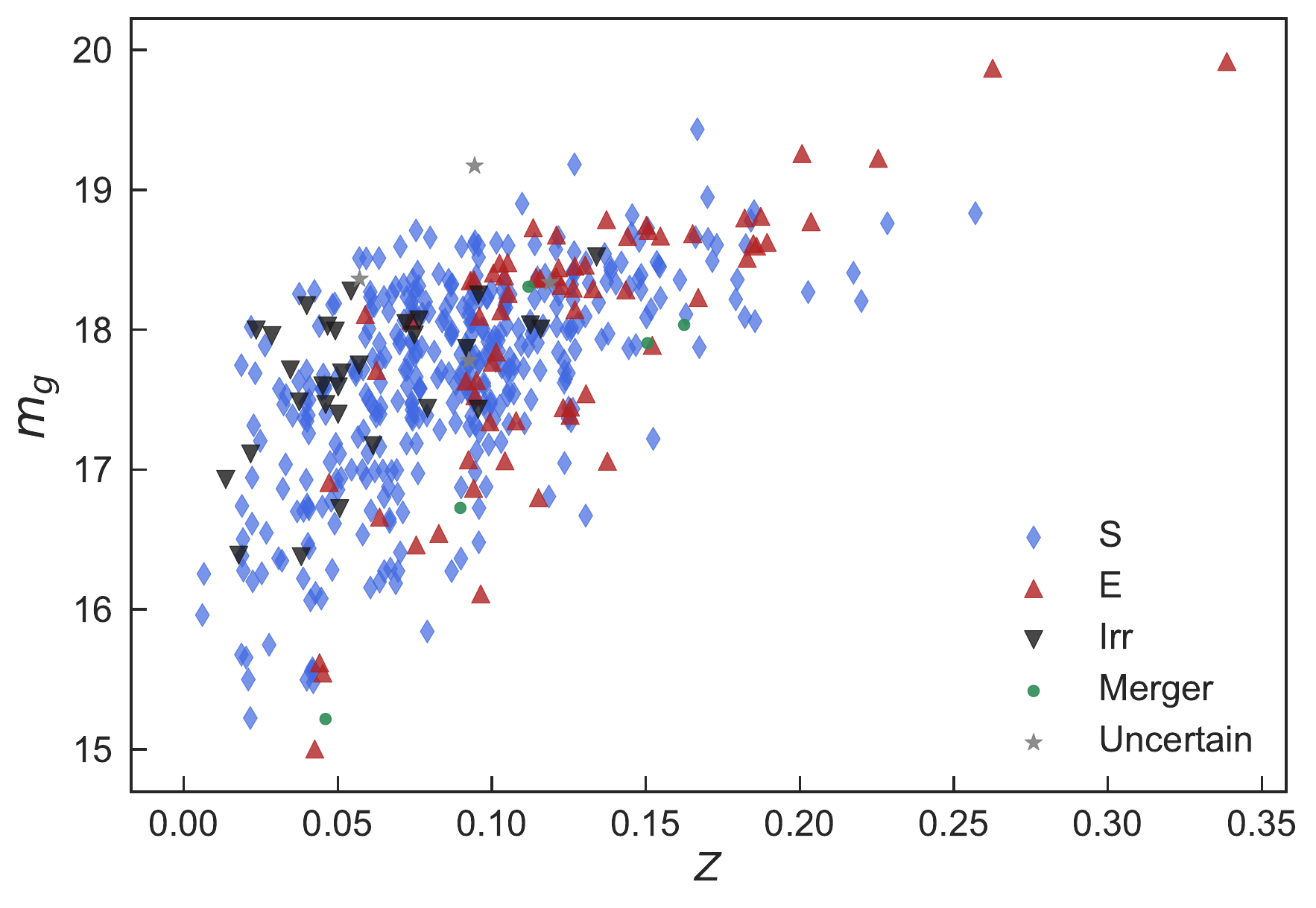}
\includegraphics[width=.45\textwidth]{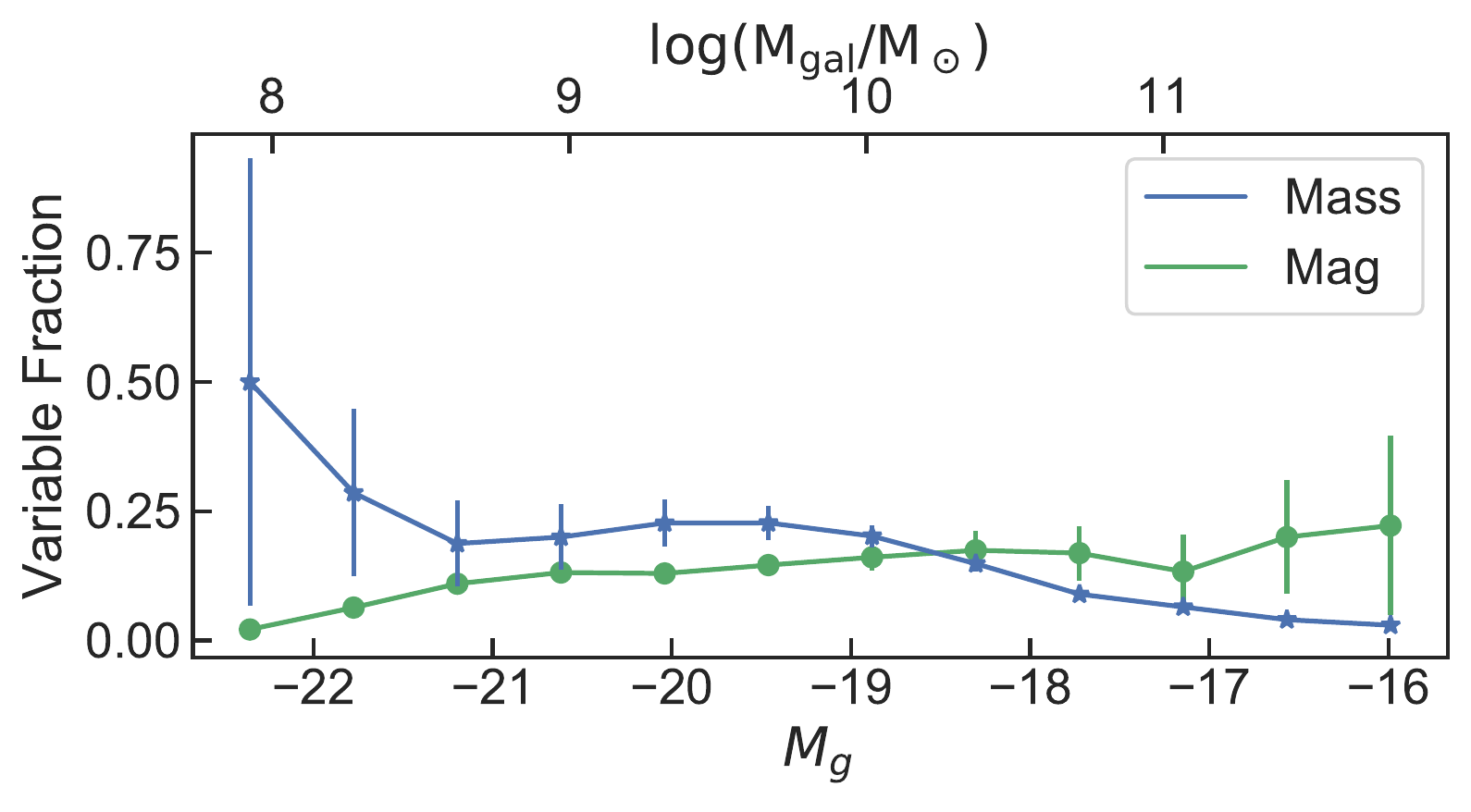}
\caption{Top: apparent \textit{g} magnitude versus redshift of the galaxy host color coded by morphological classification: spirals (S), elliptical (E), irregulars (Irr), mergers, and uncertain.
Lower: fraction of variable nuclei versus absolute \textit{g} magnitude (green line) and galaxy mass (blue line). Errors are due to Poisson statistics.
\label{fig:gma_vs_z_varGal}}
\end{figure}

Figure~\ref{fig:gma_vs_z_varGal} top panel shows the joint distribution of apparent \textit{g} band magnitude and redshift of the selected sample. Markers represent the morphological classes. 
There is a statistically significant (Anderson-Darling statistic larger than 99\% confidence value) evidence of segregation in redshift for ellipticals and irregulars, with the latter being founded at lower redshift. This is most likely driven by the fainter nature of irregulars and the difficulty to find them at large distances, and the sparsity of ellipticals in small surveyed volumes.
The largest number of candidates are spiral galaxies ($\sim\! 78\%$), which agrees with the fact that IMBHs are thought to be found in less-evolved, lower-mass galaxies. The lower panel shows the fraction of variable nuclei as a function of absolute \textit{g} band magnitude (green line) and galaxy mass (blue line) for the full galaxy sample. Errors are due to counting statistics. It can be seen that our results show a clear correlation which suggest an increase in the fraction of IMBH candidates towards smaller galaxies, being statistically significant at least for $M_g < -18$ (Kendall's $\tau = 0.89$, p-value=0.0008), and $log(M_{gal}/M_\odot) > 9.2$ (Kendall's $\tau = -0.65$, p-value=0.0095).

\subsection{X-Ray and Radio Counterparts}

We searched for X-ray and radio counterparts using the \textit{Chandra} COSMOS Survey III \citep{2012ApJS..201...30C}, the XMM-Newton Serendipitous Source Catalog Data Release 8 \citep[3XMM, ][]{2016A&A...590A...1R}, and the VLA-FIRST \citep{1994ASPC...61..165B} catalogs.

A well--established relation between the X--ray emission, radio emission and the mass of accreting BHs has been proven for a wide range of BH masses, from the stellar regime up to supermassive systems. These three quantities appear to be strongly correlated, forming what is known as the \textit{fundamental plane of black hole activity} \citep[FP, ][]{2003MNRAS.345.1057M,2004A&A...414..895F,2012MNRAS.419..267P}, which take the following form:
\begin{equation}
\log L_{R} = \xi_{RX} \log L_{X} + \xi_{RM} \log M_{BH} + b_{R},
\end{equation}
were $L_R$ is the radio Luminosity typically in the 5 GHz band, $L_X$ the X--ray luminosity in the 2-10 keV band, and $\xi_{RX}$, $\xi_{RM}$ and $b_R$ are the fitted coefficients that typically take values $0.60$, $0.78$ and $7.33$, respectively \citep{2003MNRAS.345.1057M}. The FP suggests that, at low accretion rates, the physical processes regulating the conversion of an accretion flow into radiative energy could be universal across the entire black hole mass scale. The FP is a natural consequence if black hole accretion and relativistic jet physics are scale invariants. Thus, the BH mass can be inferred from X--ray and radio luminosities. Since the FP has been populated at the low and high--end mass regime, it can be used to estimate masses for IMBH candidates. Yet the relation still suffers from very high intrinsic scatter.

21 galaxies have detected counterparts in the VLA-FIRST catalog within a 5 arcsec search radius. 
Most of these sources are found in hosts brighter than $M_g = -20$, and have a mean value of $\langle M_g\rangle \sim -20.4$. Four of them are already labeled as AGNs by the SDSS spectral classification, while 3 were labeled as BROADLINE sources in the SDSS subclass, but our spectral analysis did not detect significant broad line components (see next subsection). Nearly half of the sources (9/21) exhibit similar peak and integrated surface brightness, suggesting that the 1.4 GHz emission is coming from a compact region in the center of the galaxy.

Seven sources were found in the \textit{Chandra}-COSMOS catalog, and four in the 3XMM catalog. The hardness ratio (HR) provides a first, approximate indication of the shape of the X-ray spectra. HR is defined as $HR = (H -S)/(H + S)$, where H is the number of counts in the hard band and S is the number of counts in the soft band. Most of the sources with \textit{Chandra} counterparts are X-ray unobscured systems with ($HR$) $< -0.2$, and only one object can be classified as obscured in X-rays. No HR information is available for the 3XMM sources.

The lack of a larger number of counterparts to our IMBH candidates is not surprising.
For example, assuming a BH mass of $10^5$ M$_\odot$ accreting at $L_X = 10^{-3} L_{\textrm{Edd}}$, and using the \cite{2003MNRAS.345.1057M} FP, we expect a X-ray luminosity of $L_{X} \sim 10^{40}$ ergs s$^{-1}$ in the 2-10 keV band and $L_{R} \sim 2\times 10^{35}$ ergs s$^{-1}$ at 5 GHz. Given the distance to our candidates, typically further than 100 Mpc, the expected flux densities are $8.3 \times 10^{-15}$ ergs cm$^{-2}$ s$^{-1}$ and 0.003 mJy bm$^{-1}$ in X-ray and radio, respectively. The VLA-FIRST detection limit of 1 mJy is far above the expected radio flux density of our candidates, and the typical \textit{Chandra} exposure time of 50 ks in the COSMOS field yields to a flux limit of $2\times 10^{-15}$ ergs cm$^{-2}$ s$^{-1}$, where most of our sources were detected. Then, dedicated X-ray and radio observations are needed to prove our population of IMBH.

\subsection{BPT Diagram}

\begin{figure*}[htp]
\includegraphics[width=1.\textwidth]{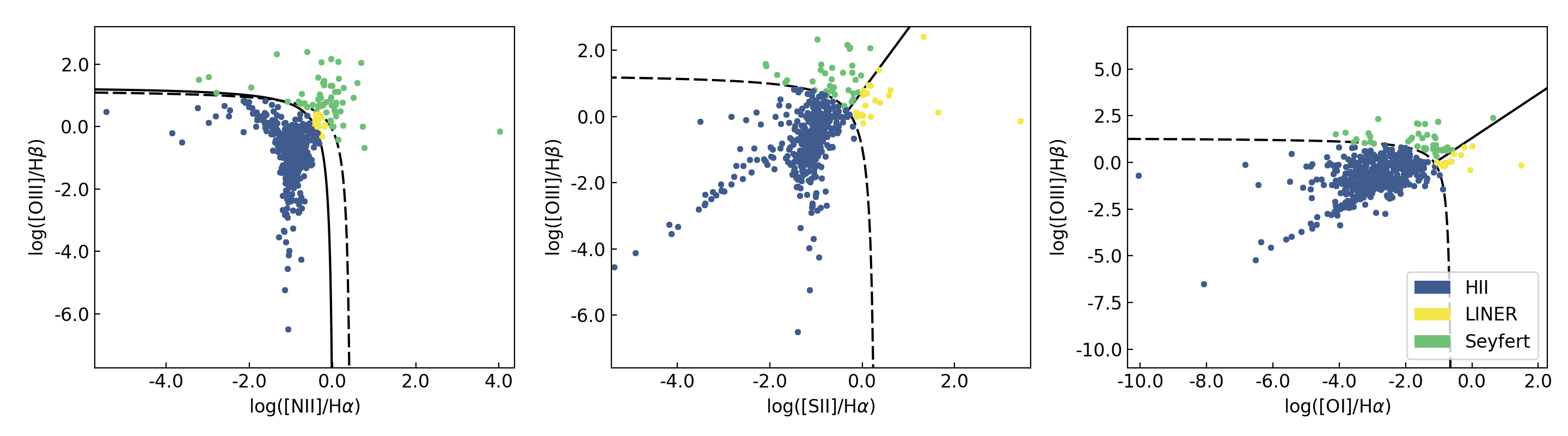}
\caption{BPT diagrams for all usable SDSS spectra using NII, SII, and OI in the left, central, and right panels. Blue dots represent star forming regions, yellow dots are LINER/Composite, and green dots are AGN/Seyfert nuclei.
\label{fig:bpt}}
\end{figure*}

We assess the possibility of the selected sample to be AGN sources by building a Baldwin, Phillips, and Terlevich diagrams \citep[BPT,][]{1981PASP...93....5B}. SDSS DR14 spectra were used for this analysis. We checked that the spectroscopic fibers were centered within 1$^{\prime \prime}$ to the centroid of the nuclear region. Therefore, the BPT diagrams match the region where variability was detected. Each spectrum was taken to the rest frame and corrected for Galactic dust reddening using the IRSA Dust Extinction Service. Spectral stellar continuum emission and absorption lines were modeled and subtracted using the STARLIGHT software \citep{2005MNRAS.358..363C}. The emission lines (namely, H$\beta$ ($\lambda$4862), OIII ($\lambda$5008), OI ($\lambda$6302), H$\alpha$ ($\lambda$6564), NII ($\lambda$6585), and SII ($\lambda$6718, $\lambda$6732)) were modeled using a Gaussian fit in order to calculate the observed flux. The final BPT diagrams are presented in Figure \ref{fig:bpt}. The different regions in the diagrams are delimited by the relations found by \cite{2006MNRAS.372..961K}, which define the following regions: AGN/Seyfert, LINER/Composite, and HII regions. 

From the 502 galaxies in our sample, 492 spectra were usable, while the rest were too noisy for spectral analysis. 22 galaxies were found in the AGN/Seyfert region of all three diagrams, and one in the LINER/Composite region. 38, 31 and 28 AGN are found by the NII, SII and OI diagrams alone, respectively. Of the 22 BPT secure sources, 9 were already labeled as AGN by SDSS. We select further 13 new secure AGNs. Combining both, our BPT diagram selection and 6 SDSS AGNs that were not simultaneously classified as AGN/Seyfert by our three BPT diagrams, 28 of our IMBH candidates can be confirmed as AGNs. Then, a $\sim5.7\%$ are AGNs among the variable galaxies (28/492), which is a larger fraction than the number of SDSS confirmed AGNs present in the variable sample $\sim3\%$ (15/502), and the fraction of AGNs in the parent galaxy sample $\sim 1.6\%$ (197/12300).

\subsection{H$\alpha$ Broad Component Analysis}

\begin{figure*}[htp]
    \gridline{\fig{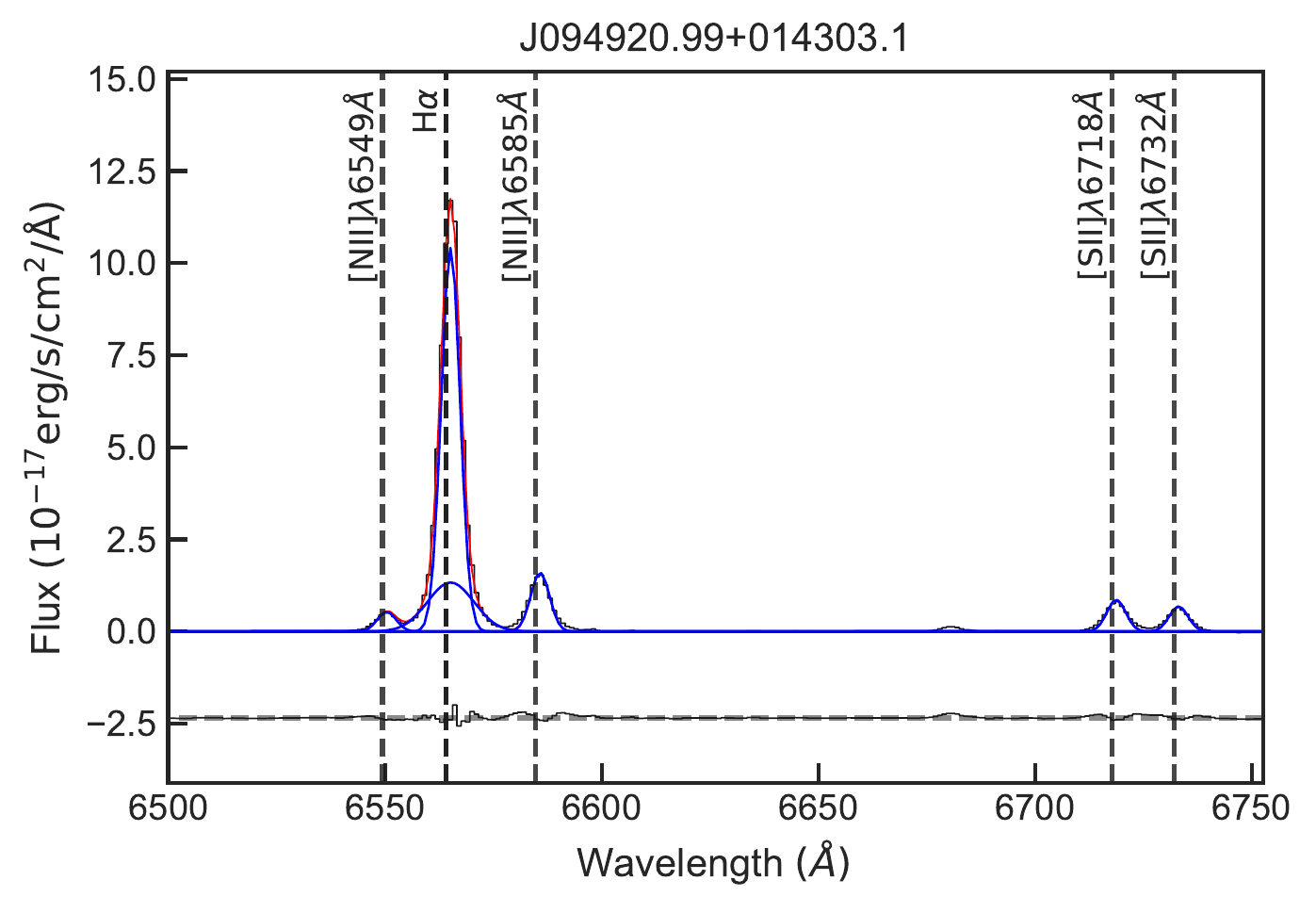}{0.33\textwidth}{(a)}
              \fig{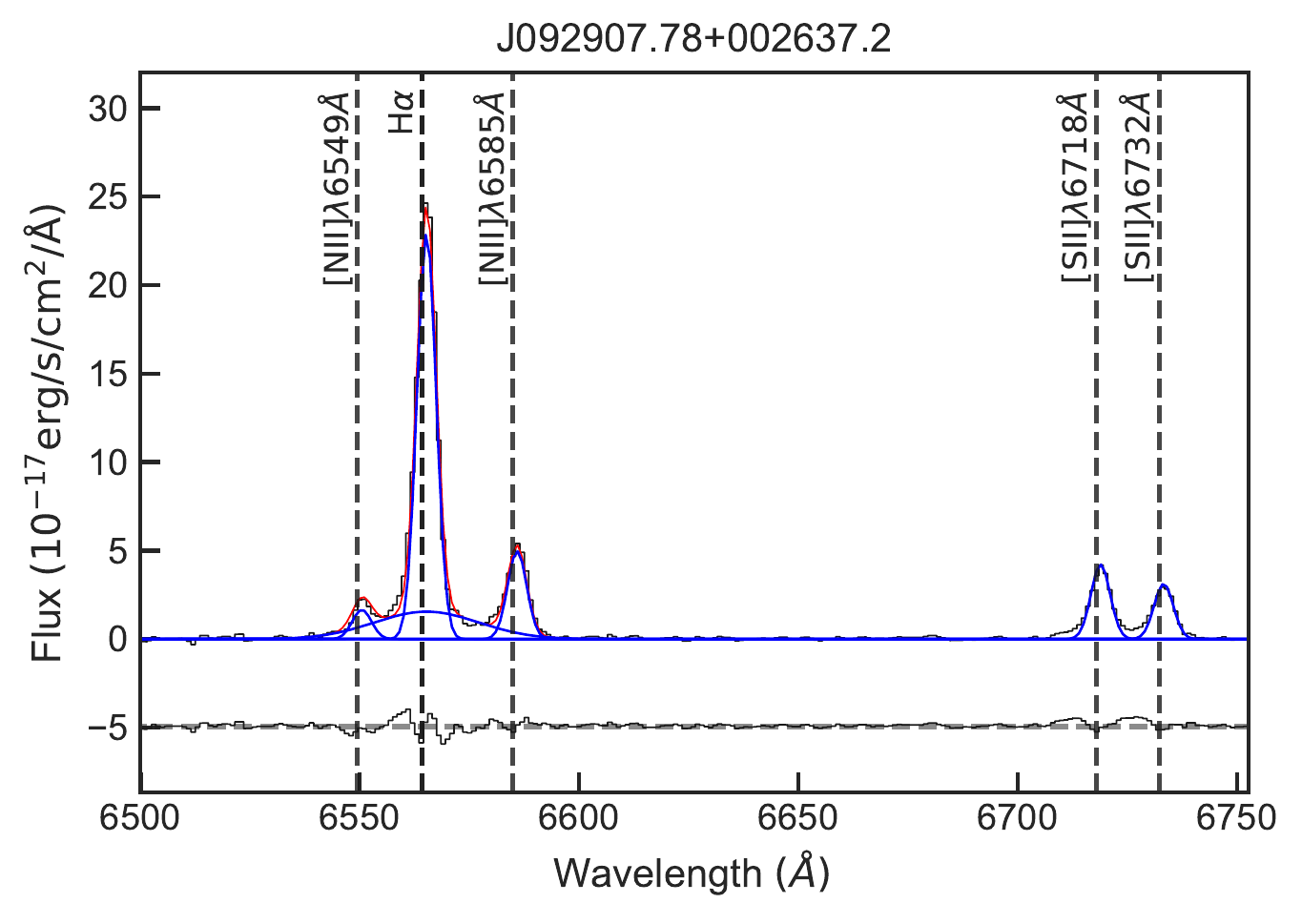}{0.33\textwidth}{(b)}
              \fig{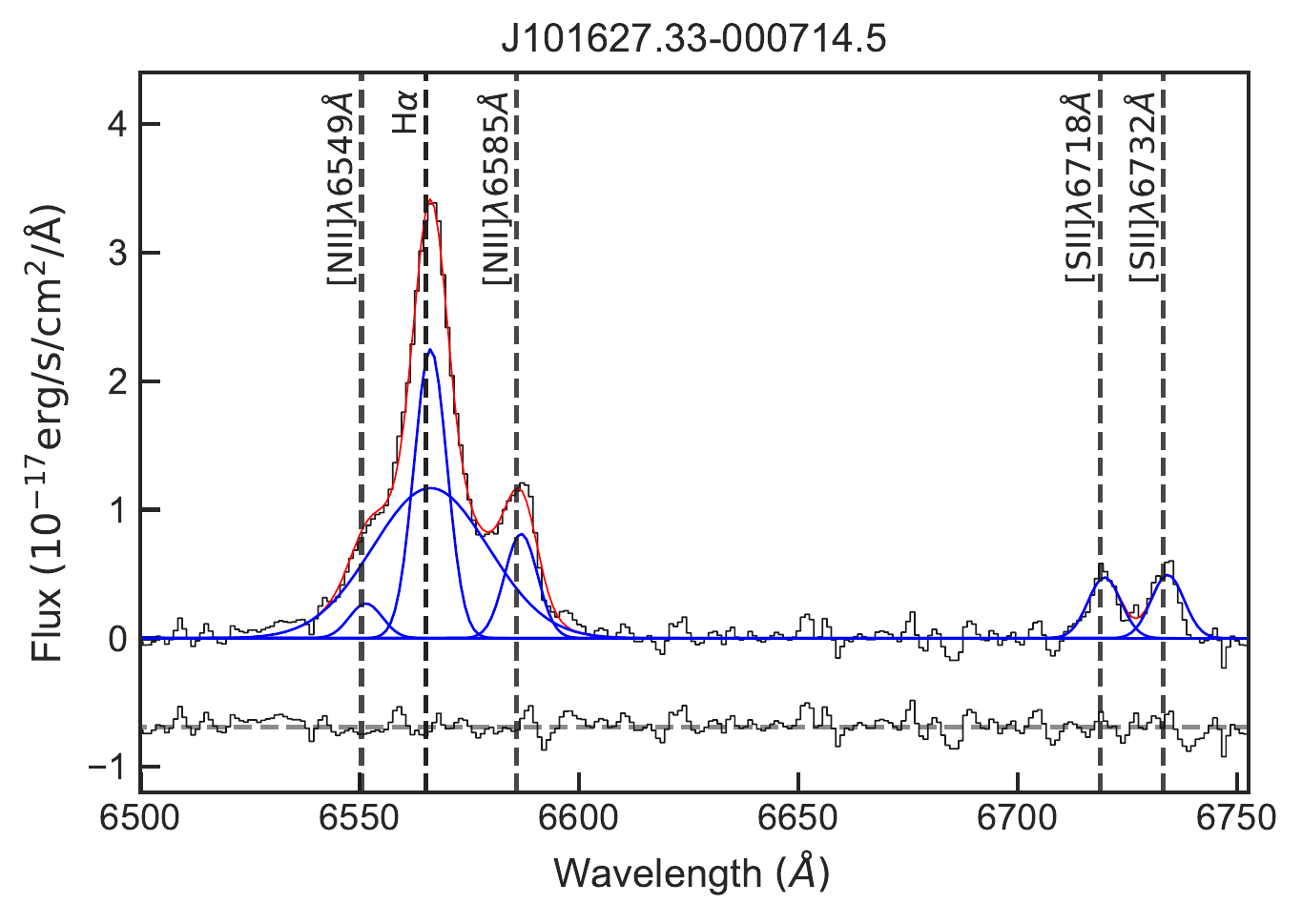}{0.33\textwidth}{(c)}}
    \caption{Spectra, in solid black line, of the three candidates with AGN confirmation from the BPT diagrams analysis. Spectral fitting shows in solid blue line the multi-components gaussians for SII ($\lambda$6718, $\lambda$6732), NII ($\lambda$6549, $\lambda$6585), and H$\alpha$ ($\lambda$6564) having narrow and broad components. Solid red line represent the addition of all components. Vertical dashed lines show the position of each components. Black solid lines at the bottom show the residuals.}
    \label{fig:spec}
\end{figure*}

We tested the existence of a broad component in the H$\alpha$ emission line in the 28 candidates confirmed as AGNs. We fitted a multi-component Gaussian profile to first the doublet SII ($\lambda$6718, $\lambda$6732), and then to the NII ($\lambda$6549, $\lambda$6585)--H$\alpha$ ($\lambda$6564) complex allowing for both narrow and broad components to H$\alpha$. Detection of a broad component with 5$\sigma$ or higher confidence were found in 3 AGNs, while the rest show no significant detection or no broad component at all. Figure \ref{fig:spec} shows the three spectral fittings. We derive BH mass estimations following the empirical correlations between broad H$\alpha$ luminosity and continuum luminosity $L_{5100}$, and the line-widths (FWHM) of H$\alpha$ and H$\beta$ \citep{2013ApJ...775..116R}:

\begin{equation}
\begin{split}
    \log\bigg(\frac{M_{BH}}{M_{\odot}}\bigg)  = &\ \log\epsilon + 6.57 \\
& +  0.47 \log \bigg( \frac{L_{\mathrm{H}\alpha}}{10^{42}\ \mathrm{erg \ s}^{-1}} \bigg) \\
& +  2.06 \log \bigg( \frac{\mathrm{FWHM}_{\mathrm{H}\alpha}}{10^3\ \mathrm{km \ s}^{-1}} \bigg)
\end{split}
\end{equation}

where $\epsilon$ is a scale factor typically having values between $\sim\!0.75\!-\!1.4$, $L_{\mathrm{H}\alpha}$ is the measured luminosity of the H$\alpha$ broad component, and FWHM$_{H\alpha}$ is its full-width-half-maximum. Here we adopt $\epsilon = 1$.

Of the three AGNs, the first is a known quasar (J094920.99+014303.1, $\log(M_{BH}/M_\odot) = 6.20 \pm 0.02$), the second is a SDSS confirmed AGN (J092907.78+002637.2, $\log(M_{BH}/M_\odot) = 6.47 \pm 0.04$), while the third is a newly classified AGN (J101627.33-000714.5, $\log(M_{BH}/M_\odot) = 6.40 \pm 0.04$). These three BHs sit just above the mass rage of IMBH and their light curves are not the most variables of our candidate sample. J101627.33-000714.5 was also selected as IMBH candidate by \cite{2018ApJS..235...40L}, our mass estimation is in agree with their estimation, $\log(M_{BH}/M_\odot) = 6.1 \pm 0.3$. Our spectral analysis was not as sensitive as those of \citeauthor{2018ApJS..235...40L} and \citeauthor{2018ApJ...863....1C}, and therefore it was expected that we might not recovered their objects.

\section{Conclusions}  \label{sec:concl}

We have introduced the SIBLING survey which selects candidate IMBH by searching for rapid variability in the nuclear region of nearby galaxies using HiTS observations. This, for the first time, used an optical, high-cadence variability selection method over a non-target dataset that covers a reasonable sky area ($168$ deg$^2$) to search for IMBHs. We have selected a sample of 502 variable galaxy nuclei over a parent sample of $\sim 12\,300$ galaxies. This sample contains local galaxies at $z \leqslant 0.35$ and covering a range of $-16 \geqslant M_g \geqslant -22$. When comparing with another systematic search of IMBH that used single-epoch spectral analysis such as that of \cite{2018ApJS..235...40L}, our selection method increases the galaxy occupancy fraction of proposed IMBH by a factor of 50 (0.04/0.0008) and the number density by a factor of 30 (3 deg$^{-2}$/ 0.1 deg$^{-2}$). Thus, confirming the presence of IMBHs in a large fraction of these candidates will dramatically increase the IMBH population. However, comparing the host properties of our candidates against the sample found in \cite{2018ApJS..235...40L} and \citet{2018ApJ...863....1C} (Figure \ref{fig:varGal_hist_papers}) a significant fraction ($30\%$) resides in smaller hosts (below $10^{10}M_{\odot}$), which seems to suggest that the same fraction of our candidates reside in the high end of the IMBH mass range, as the candidates found by \citet{2018ApJ...863....1C} Lower mass candidates would require a faster cadence search than that offered by HiTS. This is supported by the estimated characteristic time scale of the variability for a $\sim \! 10^{5}$ M$_{\odot}$ BH, of $\sim \! 0.3$ hours.

A large fraction of the candidates resides in spiral galaxies, which appears to agree with the expectation that IMBHs should be found in both less-evolve and low-mass galaxies \citep{2017IJMPD..2630021M}. The later is supported by our results as the fraction of galaxies hosting candidate IMBHs increases towards smaller galaxies. We found a small fraction of our candidates with FIRST and \textit{Chandra} detections. This is not surprising since IMBHs are supposed to be systems accreting at low rates \citep{2013ARA&A..51..511K} as the case of NGC 4395 with an accretion rate $L_{\rm bol}/L_{\rm Edd}\sim (20-2)\times 10^{-3}$ \citep{2003ApJ...588L..13F}, which together with their low mass BHs, predicts radio and X-ray fluxes of the order or below the typical detection limits of the mentioned surveys. 

Combining the BPT diagram analysis and SDSS spectral galaxy classification, 28 of our candidates are classified as AGN. Three of them having detectable H$\alpha$ broad component, which lead to BH mass estimations of $\sim\! 10^6 M_\odot$. The lack of a positive classification in the remaining sources might be due to contamination from nearby, bright star-forming regions, stellar dilution, or to changes in the line ratios in the IMBH regime, as predicted by \cite{2015ApJ...811...26T}. In fact, optical spectroscopic diagnostics such as the BPT diagrams are expected to fail for low-mass BHs \citep{2019ApJ...870L...2C}.

\acknowledgments

J.M.~acknowledges the support from CONICYT Chile through (PCHA/Doctorado-Nacional/2014-21140892). 
J.M.~and F.F.~acknowledge support from the Ministry of Economy, Development, and Tourism's Millennium Science Initiative through grant IC120009, awarded to the Millennium Institute of Astrophysics.
P.L. acknowledges support by Fondecyt through Project No.~1161184.
F.F.~acknowledges support by Fondecyt through Project No.~11130228.
J.M. and F.F.~acknowledge support from Basal Project PFB-03, Centro de Modelamiento Matemáico, Universidad de Chile.
Powered@NLHPC: this research was supported by the High Performance Computing infrastructure of the National Laboratory for High Performance Computing (NLHPC), PIA ECM-02, CONICYT.
This project used data obtained with the Dark Energy Camera (DECam),
which was constructed by the Dark Energy Survey (DES) collaboration.
Funding for the DES Projects has been provided by 
the U.S. Department of Energy, 
the U.S. National Science Foundation, 
the Ministry of Science and Education of Spain, 
the Science and Technology Facilities Council of the United Kingdom, 
the Higher Education Funding Council for England, 
the National Center for Supercomputing Applications at the University of Illinois at Urbana-Champaign, 
the Kavli Institute of Cosmological Physics at the University of Chicago, 
the Center for Cosmology and Astro-Particle Physics at the Ohio State University, 
the Mitchell Institute for Fundamental Physics and Astronomy at Texas A\&M University, 
Financiadora de Estudos e Projetos, Funda{\c c}{\~a}o Carlos Chagas Filho de Amparo {\`a} Pesquisa do Estado do Rio de Janeiro, 
Conselho Nacional de Desenvolvimento Cient{\'i}fico e Tecnol{\'o}gico and the Minist{\'e}rio da Ci{\^e}ncia, Tecnologia e Inovac{\~a}o, 
the Deutsche Forschungsgemeinschaft, 
and the Collaborating Institutions in the Dark Energy Survey. 
The Collaborating Institutions are 
Argonne National Laboratory, 
the University of California at Santa Cruz, 
the University of Cambridge, 
Centro de Investigaciones En{\'e}rgeticas, Medioambientales y Tecnol{\'o}gicas-Madrid, 
the University of Chicago, 
University College London, 
the DES-Brazil Consortium, 
the University of Edinburgh, 
the Eidgen{\"o}ssische Technische Hoch\-schule (ETH) Z{\"u}rich, 
Fermi National Accelerator Laboratory, 
the University of Illinois at Urbana-Champaign, 
the Institut de Ci{\`e}ncies de l'Espai (IEEC/CSIC), 
the Institut de F{\'i}sica d'Altes Energies, 
Lawrence Berkeley National Laboratory, 
the Ludwig-Maximilians Universit{\"a}t M{\"u}nchen and the associated Excellence Cluster Universe, 
the University of Michigan, 
{the} National Optical Astronomy Observatory, 
the University of Nottingham, 
the Ohio State University, 
the OzDES Membership Consortium
the University of Pennsylvania, 
the University of Portsmouth, 
SLAC National Accelerator Laboratory, 
Stanford University, 
the University of Sussex, 
and Texas A\&M University.
Based on observations at Cerro Tololo Inter-American Observatory, National Optical Astronomy Observatory (Prop. 2015A/14A-0608 and F. Förster), which is operated by the Association of Universities for Research in Astronomy (AURA) under a cooperative agreement with the National Science Foundation. 

\vspace{5mm}
\facilities{CTIO:4m, DECam}

\bibliography{bibliography.bib}

\begin{thebibliography}{}
\expandafter\ifx\csname natexlab\endcsname\relax\def\natexlab#1{#1}\fi
\providecommand{\url}[1]{\href{#1}{#1}}

\bibitem[{Abbott {et~al.}(2016)Abbott, Abbott, Abbott, Abernathy, Acernese,
  Ackley, Adams, Adams, Addesso, Adhikari, Adya, Affeldt, Agathos, Agatsuma,
  Aggarwal, Aguiar, Aiello, Ain, Ajith, Allen, Allocca, Altin, Anderson,
  Anderson, Arai, Arain, Araya, Arceneaux, Areeda, Arnaud, Arun, Ascenzi,
  Ashton, Ast, Aston, Astone, Aufmuth, Aulbert, Babak, Bacon, Bader, Baker,
  Baldaccini, Ballardin, Ballmer, Barayoga, Barclay, Barish, Barker, Barone,
  Barr, Barsotti, Barsuglia, Barta, Bartlett, Barton, Bartos, Bassiri, Basti,
  Batch, Baune, Bavigadda, Bazzan, Behnke, Bejger, Belczynski, Bell, Bell,
  Berger, Bergman, Bergmann, Berry, Bersanetti, Bertolini, Betzwieser, Bhagwat,
  Bhandare, Bilenko, Billingsley, Birch, Birney, Birnholtz, Biscans, Bisht,
  Bitossi, Biwer, Bizouard, Blackburn, Blair, Blair, Blair, Bloemen, Bock,
  Bodiya, Boer, Bogaert, Bogan, Bohe, Bojtos, Bond, Bondu, Bonnand, Boom, Bork,
  Boschi, Bose, Bouffanais, Bozzi, Bradaschia, Brady, Braginsky, Branchesi,
  Brau, Briant, Brillet, Brinkmann, Brisson, Brockill, Brooks, Brown, Brown,
  Brown, Buchanan, Buikema, Bulik, Bulten, Buonanno, Buskulic, Buy, Byer,
  Cabero, Cadonati, Cagnoli, Cahillane, Bustillo, Callister, Calloni, Camp,
  Cannon, Cao, Capano, Capocasa, Carbognani, Caride, Diaz, Casentini, Caudill,
  Cavagli\`a, Cavalier, Cavalieri, Cella, Cepeda, Baiardi, Cerretani, Cesarini,
  Chakraborty, Chalermsongsak, Chamberlin, Chan, Chao, Charlton,
  Chassande-Mottin, Chen, Chen, Cheng, Chincarini, Chiummo, Cho, Cho, Chow,
  Christensen, Chu, Chua, Chung, Ciani, Clara, Clark, Cleva, Coccia, Cohadon,
  Colla, Collette, Cominsky, Constancio, Conte, Conti, Cook, Corbitt, Cornish,
  Corsi, Cortese, Costa, Coughlin, Coughlin, Coulon, Countryman, Couvares,
  Cowan, Coward, Cowart, Coyne, Coyne, Craig, Creighton, Creighton, Cripe,
  Crowder, Cruise, Cumming, Cunningham, Cuoco, Canton, Danilishin, D'Antonio,
  Danzmann, Darman, Da~Silva~Costa, Dattilo, Dave, Daveloza, Davier, Davies,
  Daw, Day, De, DeBra, Debreczeni, Degallaix, De~Laurentis, Del\'eglise,
  Del~Pozzo, Denker, Dent, Dereli, Dergachev, DeRosa, De~Rosa, DeSalvo,
  Dhurandhar, D\'{\i}az, Di~Fiore, Di~Giovanni, Di~Lieto, Di~Pace, Di~Palma,
  Di~Virgilio, Dojcinoski, Dolique, Donovan, Dooley, Doravari, Douglas, Downes,
  Drago, Drever, Driggers, Du, Ducrot, Dwyer, Edo, Edwards, Effler, Eggenstein,
  Ehrens, Eichholz, Eikenberry, Engels, Essick, Etzel, Evans, Evans, Everett,
  Factourovich, Fafone, Fair, Fairhurst, Fan, Fang, Farinon, Farr, Farr,
  Favata, Fays, Fehrmann, Fejer, Feldbaum, Ferrante, Ferreira, Ferrini,
  Fidecaro, Finn, Fiori, Fiorucci, Fisher, Flaminio, Fletcher, Fong, Fournier,
  Franco, Frasca, Frasconi, Frede, Frei, Freise, Frey, Frey, Fricke, Fritschel,
  Frolov, Fulda, Fyffe, Gabbard, Gair, Gammaitoni, Gaonkar, Garufi, Gatto,
  Gaur, Gehrels, Gemme, Gendre, Genin, Gennai, George, Gergely, Germain, Ghosh,
  Ghosh, Ghosh, Giaime, Giardina, Giazotto, Gill, Glaefke, Gleason, Goetz,
  Goetz, Gondan, Gonz\'alez, Castro, Gopakumar, Gordon, Gorodetsky, Gossan,
  Gosselin, Gouaty, Graef, Graff, Granata, Grant, Gras, Gray, Greco, Green,
  Greenhalgh, Groot, Grote, Grunewald, Guidi, Guo, Gupta, Gupta, Gushwa,
  Gustafson, Gustafson, Hacker, Hall, Hall, Hammond, Haney, Hanke, Hanks,
  Hanna, Hannam, Hanson, Hardwick, Harms, Harry, Harry, Hart, Hartman, Haster,
  Haughian, Healy, Heefner, Heidmann, Heintze, Heinzel, Heitmann, Hello,
  Hemming, Hendry, Heng, Hennig, Heptonstall, Heurs, Hild, Hoak, Hodge, Hofman,
  Hollitt, Holt, Holz, Hopkins, Hosken, Hough, Houston, Howell, Hu, Huang,
  Huerta, Huet, Hughey, Husa, Huttner, Huynh-Dinh, Idrisy, Indik, Ingram, Inta,
  Isa, Isac, Isi, Islas, Isogai, Iyer, Izumi, Jacobson, Jacqmin, Jang, Jani,
  Jaranowski, Jawahar, Jim\'enez-Forteza, Johnson, Johnson-McDaniel, Jones,
  Jones, Jonker, Ju, Haris, Kalaghatgi, Kalogera, Kandhasamy, Kang, Kanner,
  Karki, Kasprzack, Katsavounidis, Katzman, Kaufer, Kaur, Kawabe, Kawazoe,
  K\'ef\'elian, Kehl, Keitel, Kelley, Kells, Kennedy, Keppel, Key,
  Khalaidovski, Khalili, Khan, Khan, Khan, Khazanov, Kijbunchoo, Kim, Kim, Kim,
  Kim, Kim, Kim, King, King, Kinzel, Kissel, Kleybolte, Klimenko, Koehlenbeck,
  Kokeyama, Koley, Kondrashov, Kontos, Koranda, Korobko, Korth, Kowalska,
  Kozak, Kringel, Krishnan, Kr\'olak, Krueger, Kuehn, Kumar, Kumar, Kuo,
  Kutynia, Kwee, Lackey, Landry, Lange, Lantz, Lasky, Lazzarini, Lazzaro,
  Leaci, Leavey, Lebigot, Lee, Lee, Lee, Lee, Lenon, Leonardi, Leong, Leroy,
  Letendre, Levin, Levine, Li, Libson, Littenberg, Lockerbie, Logue, Lombardi,
  London, Lord, Lorenzini, Loriette, Lormand, Losurdo, Lough, Lousto, Lovelace,
  L\"uck, Lundgren, Luo, Lynch, Ma, MacDonald, Machenschalk, MacInnis, Macleod,
  Maga\~na Sandoval, Magee, Mageswaran, Majorana, Maksimovic, Malvezzi, Man,
  Mandel, Mandic, Mangano, Mansell, Manske, Mantovani, Marchesoni, Marion,
  M\'arka, M\'arka, Markosyan, Maros, Martelli, Martellini, Martin, Martin,
  Martynov, Marx, Mason, Masserot, Massinger, Masso-Reid, Matichard, Matone,
  Mavalvala, Mazumder, Mazzolo, McCarthy, McClelland, McCormick, McGuire,
  McIntyre, McIver, McManus, McWilliams, Meacher, Meadors, Meidam, Melatos,
  Mendell, Mendoza-Gandara, Mercer, Merilh, Merzougui, Meshkov, Messenger,
  Messick, Meyers, Mezzani, Miao, Michel, Middleton, Mikhailov, Milano, Miller,
  Millhouse, Minenkov, Ming, Mirshekari, Mishra, Mitra, Mitrofanov,
  Mitselmakher, Mittleman, Moggi, Mohan, Mohapatra, Montani, Moore, Moore,
  Moraru, Moreno, Morriss, Mossavi, Mours, Mow-Lowry, Mueller, Mueller, Muir,
  Mukherjee, Mukherjee, Mukherjee, Mukund, Mullavey, Munch, Murphy, Murray,
  Mytidis, Nardecchia, Naticchioni, Nayak, Necula, Nedkova, Nelemans, Neri,
  Neunzert, Newton, Nguyen, Nielsen, Nissanke, Nitz, Nocera, Nolting,
  Normandin, Nuttall, Oberling, Ochsner, O'Dell, Oelker, Ogin, Oh, Oh, Ohme,
  Oliver, Oppermann, Oram, O'Reilly, O'Shaughnessy, Ott, Ottaway, Ottens,
  Overmier, Owen, Pai, Pai, Palamos, Palashov, Palomba, Pal-Singh, Pan, Pan,
  Pankow, Pannarale, Pant, Paoletti, Paoli, Papa, Paris, Parker, Pascucci,
  Pasqualetti, Passaquieti, Passuello, Patricelli, Patrick, Pearlstone,
  Pedraza, Pedurand, Pekowsky, Pele, Penn, Perreca, Pfeiffer, Phelps, Piccinni,
  Pichot, Pickenpack, Piergiovanni, Pierro, Pillant, Pinard, Pinto, Pitkin,
  Poeld, Poggiani, Popolizio, Post, Powell, Prasad, Predoi, Premachandra,
  Prestegard, Price, Prijatelj, Principe, Privitera, Prix, Prodi, Prokhorov,
  Puncken, Punturo, Puppo, P\"urrer, Qi, Qin, Quetschke, Quintero,
  Quitzow-James, Raab, Rabeling, Radkins, Raffai, Raja, Rakhmanov, Ramet,
  Rapagnani, Raymond, Razzano, Re, Read, Reed, Regimbau, Rei, Reid, Reitze,
  Rew, Reyes, Ricci, Riles, Robertson, Robie, Robinet, Rocchi, Rolland,
  Rollins, Roma, Romano, Romano, Romanov, Romie, Rosi\ifmmode~\acute{n}\else
  \'{n}\fi{}ska, Rowan, R\"udiger, Ruggi, Ryan, Sachdev, Sadecki, Sadeghian,
  Salconi, Saleem, Salemi, Samajdar, Sammut, Sampson, Sanchez, Sandberg,
  Sandeen, Sanders, Sanders, Sassolas, Sathyaprakash, Saulson, Sauter, Savage,
  Sawadsky, Schale, Schilling, Schmidt, Schmidt, Schnabel, Schofield,
  Sch\"onbeck, Schreiber, Schuette, Schutz, Scott, Scott, Sellers, Sengupta,
  Sentenac, Sequino, Sergeev, Serna, Setyawati, Sevigny, Shaddock, Shaffer,
  Shah, Shahriar, Shaltev, Shao, Shapiro, Shawhan, Sheperd, Shoemaker,
  Shoemaker, Siellez, Siemens, Sigg, Silva, Simakov, Singer, Singer, Singh,
  Singh, Singhal, Sintes, Slagmolen, Smith, Smith, Smith, Smith, Son, Sorazu,
  Sorrentino, Souradeep, Srivastava, Staley, Steinke, Steinlechner,
  Steinlechner, Steinmeyer, Stephens, Stevenson, Stone, Strain, Straniero,
  Stratta, Strauss, Strigin, Sturani, Stuver, Summerscales, Sun, Sutton,
  Swinkels, Szczepa\ifmmode~\acute{n}\else \'{n}\fi{}czyk, Tacca, Talukder,
  Tanner, T\'apai, Tarabrin, Taracchini, Taylor, Theeg, Thirugnanasambandam,
  Thomas, Thomas, Thomas, Thorne, Thorne, Thrane, Tiwari, Tiwari, Tokmakov,
  Tomlinson, Tonelli, Torres, Torrie, T\"oyr\"a, Travasso, Traylor, Trifir\`o,
  Tringali, Trozzo, Tse, Turconi, Tuyenbayev, Ugolini, Unnikrishnan, Urban,
  Usman, Vahlbruch, Vajente, Valdes, Vallisneri, van Bakel, van Beuzekom,
  van~den Brand, Van Den~Broeck, Vander-Hyde, van~der Schaaf, van Heijningen,
  van Veggel, Vardaro, Vass, Vas\'uth, Vaulin, Vecchio, Vedovato, Veitch,
  Veitch, Venkateswara, Verkindt, Vetrano, Vicer\'e, Vinciguerra, Vine, Vinet,
  Vitale, Vo, Vocca, Vorvick, Voss, Vousden, Vyatchanin, Wade, Wade, Wade,
  Waldman, Walker, Wallace, Walsh, Wang, Wang, Wang, Wang, Wang, Ward, Ward,
  Warner, Was, Weaver, Wei, Weinert, Weinstein, Weiss, Welborn, Wen,
  We\ss{}els, Westphal, Wette, Whelan, Whitcomb, White, Whiting, Wiesner,
  Wilkinson, Willems, Williams, Williams, Williamson, Willis, Willke, Wimmer,
  Winkelmann, Winkler, Wipf, Wiseman, Wittel, Woan, Worden, Wright, Wu, Yablon,
  Yakushin, Yam, Yamamoto, Yancey, Yap, Yu, Yvert, Zadro\ifmmode~\dot{z}\else
  \.{z}\fi{}ny, Zangrando, Zanolin, Zendri, Zevin, Zhang, Zhang, Zhang, Zhang,
  Zhao, Zhou, Zhou, Zhu, Zucker, Zuraw, \& Zweizig}]{PhysRevLett.116.061102}
Abbott, B.~P., Abbott, R., Abbott, T.~D., {et~al.} 2016, Phys. Rev. Lett., 116,
  061102.
\newblock \url{https://link.aps.org/doi/10.1103/PhysRevLett.116.061102}

\bibitem[{Abbott {et~al.}(2017)Abbott, Abbott, Abbott, Acernese, Ackley, Adams,
  Adams, Addesso, Adhikari, Adya, Affeldt, Afrough, Agarwal, Agathos, Agatsuma,
  Aggarwal, Aguiar, Aiello, Ain, Ajith, Allen, Allen, Allocca, Altin, Amato,
  Ananyeva, Anderson, Anderson, Antier, Appert, Arai, Araya, Areeda, Arnaud,
  Arun, Ascenzi, Ashton, Ast, Aston, Astone, Aufmuth, Aulbert, AultONeal,
  Avila-Alvarez, Babak, Bacon, Bader, Bae, Baker, Baldaccini, Ballardin,
  Ballmer, Banagiri, Barayoga, Barclay, Barish, Barker, Barone, Barr, Barsotti,
  Barsuglia, Barta, Bartlett, Bartos, Bassiri, Basti, Batch, Baune, Bawaj,
  Bazzan, B\'ecsy, Beer, Bejger, Belahcene, Bell, Berger, Bergmann, Berry,
  Bersanetti, Bertolini, Betzwieser, Bhagwat, Bhandare, Bilenko, Billingsley,
  Billman, Birch, Birney, Birnholtz, Biscans, Bisht, Bitossi, Biwer, Bizouard,
  Blackburn, Blackman, Blair, Blair, Blair, Bloemen, Bock, Bode, Boer, Bogaert,
  Bohe, Bondu, Bonnand, Boom, Bork, Boschi, Bose, Bouffanais, Bozzi,
  Bradaschia, Brady, Braginsky, Branchesi, Brau, Briant, Brillet, Brinkmann,
  Brisson, Brockill, Broida, Brooks, Brown, Brown, Brown, Brunett, Buchanan,
  Buikema, Bulik, Bulten, Buonanno, Buskulic, Buy, Byer, Cabero, Cadonati,
  Cagnoli, Cahillane, Calder\'on~Bustillo, Callister, Calloni, Camp, Canepa,
  Canizares, Cannon, Cao, Cao, Capano, Capocasa, Carbognani, Caride, Carney,
  Casanueva~Diaz, Casentini, Caudill, Cavagli\`a, Cavalier, Cavalieri, Cella,
  Cepeda, Cerboni~Baiardi, Cerretani, Cesarini, Chamberlin, Chan, Chao,
  Charlton, Chassande-Mottin, Chatterjee, Chatziioannou, Cheeseboro, Chen,
  Chen, Cheng, Chincarini, Chiummo, Chmiel, Cho, Cho, Chow, Christensen, Chu,
  Chua, Chua, Chung, Chung, Ciani, Ciolfi, Cirelli, Cirone, Clara, Clark,
  Cleva, Cocchieri, Coccia, Cohadon, Colla, Collette, Cominsky, Constancio,
  Conti, Cooper, Corban, Corbitt, Corley, Cornish, Corsi, Cortese, Costa,
  Coughlin, Coughlin, Coulon, Countryman, Couvares, Covas, Cowan, Coward,
  Cowart, Coyne, Coyne, Creighton, Creighton, Cripe, Crowder, Cullen, Cumming,
  Cunningham, Cuoco, Dal~Canton, Danilishin, D'Antonio, Danzmann, Dasgupta,
  Da~Silva~Costa, Dattilo, Dave, Davier, Davis, Daw, Day, De, DeBra, Deelman,
  Degallaix, De~Laurentis, Del\'eglise, Del~Pozzo, Denker, Dent, Dergachev,
  De~Rosa, DeRosa, DeSalvo, Devenson, Devine, Dhurandhar, D\'{\i}az, Di~Fiore,
  Di~Giovanni, Di~Girolamo, Di~Lieto, Di~Pace, Di~Palma, Di~Renzo, Doctor,
  Dolique, Donovan, Dooley, Doravari, Dorrington, Douglas, Dovale~\'Alvarez,
  Downes, Drago, Drever, Driggers, Du, Ducrot, Duncan, Dwyer, Edo, Edwards,
  Effler, Eggenstein, Ehrens, Eichholz, Eikenberry, Eisenstein, Essick,
  Etienne, Etzel, Evans, Evans, Factourovich, Fafone, Fair, Fairhurst, Fan,
  Farinon, Farr, Farr, Fauchon-Jones, Favata, Fays, Fehrmann, Feicht, Fejer,
  Fernandez-Galiana, Ferrante, Ferreira, Ferrini, Fidecaro, Fiori, Fiorucci,
  Fisher, Flaminio, Fletcher, Fong, Forsyth, Forsyth, Fournier, Frasca,
  Frasconi, Frei, Freise, Frey, Frey, Fries, Fritschel, Frolov, Fulda, Fyffe,
  Gabbard, Gabel, Gadre, Gaebel, Gair, Gammaitoni, Ganija, Gaonkar, Garufi,
  Gaudio, Gaur, Gayathri, Gehrels, Gemme, Genin, Gennai, George, George,
  Gergely, Germain, Ghonge, Ghosh, Ghosh, Ghosh, Giaime, Giardina, Giazotto,
  Gill, Glover, Goetz, Goetz, Gomes, Gonz\'alez, Gonzalez~Castro, Gopakumar,
  Gorodetsky, Gossan, Gosselin, Gouaty, Grado, Graef, Granata, Grant, Gras,
  Gray, Greco, Green, Groot, Grote, Grunewald, Gruning, Guidi, Guo, Gupta,
  Gupta, Gushwa, Gustafson, Gustafson, Hall, Hall, Hammond, Haney, Hanke,
  Hanks, Hanna, Hannam, Hannuksela, Hanson, Hardwick, Harms, Harry, Harry,
  Hart, Haster, Haughian, Healy, Heidmann, Heintze, Heitmann, Hello, Hemming,
  Hendry, Heng, Hennig, Henry, Heptonstall, Heurs, Hild, Hoak, Hofman, Holt,
  Holz, Hopkins, Horst, Hough, Houston, Howell, Hu, Huerta, Huet, Hughey, Husa,
  Huttner, Huynh-Dinh, Indik, Ingram, Inta, Intini, Isa, Isac, Isi, Iyer,
  Izumi, Jacqmin, Jani, Jaranowski, Jawahar, Jim\'enez-Forteza, Johnson,
  Johnson-McDaniel, Jones, Jones, Jonker, Ju, Junker, Kalaghatgi, Kalogera,
  Kandhasamy, Kang, Kanner, Karki, Karvinen, Kasprzack, Katolik, Katsavounidis,
  Katzman, Kaufer, Kawabe, K\'ef\'elian, Keitel, Kemball, Kennedy, Kent, Key,
  Khalili, Khan, Khan, Khan, Khazanov, Kijbunchoo, Kim, Kim, Kim, Kim, Kim,
  Kimbrell, King, King, Kirchhoff, Kissel, Kleybolte, Klimenko, Koch,
  Koehlenbeck, Koley, Kondrashov, Kontos, Korobko, Korth, Kowalska, Kozak,
  Kr\"amer, Kringel, Krishnan, Kr\'olak, Kuehn, Kumar, Kumar, Kumar, Kuo,
  Kutynia, Kwang, Lackey, Lai, Landry, Lang, Lange, Lantz, Lanza,
  Lartaux-Vollard, Lasky, Laxen, Lazzarini, Lazzaro, Leaci, Leavey, Lee, Lee,
  Lee, Lee, Lee, Lehmann, Lenon, Leonardi, Leroy, Letendre, Levin, Li, Libson,
  Littenberg, Liu, Lo, Lockerbie, London, Lord, Lorenzini, Loriette, Lormand,
  Losurdo, Lough, Lovelace, L\"uck, Lumaca, Lundgren, Lynch, Ma, Macfoy,
  Machenschalk, MacInnis, Macleod, Maga\~na Hernandez, Maga\~na Sandoval,
  Maga\~na Zertuche, Magee, Majorana, Maksimovic, Man, Mandic, Mangano,
  Mansell, Manske, Mantovani, Marchesoni, Marion, M\'arka, M\'arka, Markakis,
  Markosyan, Maros, Martelli, Martellini, Martin, Martynov, Marx, Mason,
  Masserot, Massinger, Masso-Reid, Mastrogiovanni, Matas, Matichard, Matone,
  Mavalvala, Mayani, Mazumder, McCarthy, McClelland, McCormick, McCuller,
  McGuire, McIntyre, McIver, McManus, McRae, McWilliams, Meacher, Meadors,
  Meidam, Mejuto-Villa, Melatos, Mendell, Mercer, Merilh, Merzougui, Meshkov,
  Messenger, Messick, Metzdorff, Meyers, Mezzani, Miao, Michel, Middleton,
  Mikhailov, Milano, Miller, Miller, Miller, Miller, Millhouse, Minazzoli,
  Minenkov, Ming, Mishra, Mitra, Mitrofanov, Mitselmakher, Mittleman, Moggi,
  Mohan, Mohapatra, Montani, Moore, Moore, Moraru, Moreno, Morriss, Mours,
  Mow-Lowry, Mueller, Muir, Mukherjee, Mukherjee, Mukherjee, Mukund, Mullavey,
  Munch, Muniz, Murray, Napier, Nardecchia, Naticchioni, Nayak, Nelemans,
  Nelson, Neri, Nery, Neunzert, Newport, Newton, Ng, Nguyen, Nichols, Nielsen,
  Nissanke, Nitz, Noack, Nocera, Nolting, Normandin, Nuttall, Oberling,
  Ochsner, Oelker, Ogin, Oh, Oh, Ohme, Oliver, Oppermann, Oram, O'Reilly,
  Ormiston, Ortega, O'Shaughnessy, Ottaway, Overmier, Owen, Pace, Page, Page,
  Pai, Pai, Palamos, Palashov, Palomba, Pal-Singh, Pan, Pang, Pang, Pankow,
  Pannarale, Pant, Paoletti, Paoli, Papa, Paris, Parker, Pascucci, Pasqualetti,
  Passaquieti, Passuello, Patricelli, Pearlstone, Pedraza, Pedurand, Pekowsky,
  Pele, Penn, Perez, Perreca, Perri, Pfeiffer, Phelps, Piccinni, Pichot,
  Piergiovanni, Pierro, Pillant, Pinard, Pinto, Pitkin, Poggiani, Popolizio,
  Porter, Post, Powell, Prasad, Pratt, Predoi, Prestegard, Prijatelj, Principe,
  Privitera, Prodi, Prokhorov, Puncken, Punturo, Puppo, P\"urrer, Qi, Qin, Qiu,
  Quetschke, Quintero, Quitzow-James, Raab, Rabeling, Radkins, Raffai, Raja,
  Rajan, Rakhmanov, Ramirez, Rapagnani, Raymond, Razzano, Read, Regimbau, Rei,
  Reid, Reitze, Rew, Reyes, Ricci, Ricker, Rieger, Riles, Rizzo, Robertson,
  Robie, Robinet, Rocchi, Rolland, Rollins, Roma, Romano, Romano, Romel, Romie,
  Rosi\ifmmode~\acute{n}\else \'{n}\fi{}ska, Ross, Rowan, R\"udiger, Ruggi,
  Ryan, Rynge, Sachdev, Sadecki, Sadeghian, Sakellariadou, Salconi, Saleem,
  Salemi, Samajdar, Sammut, Sampson, Sanchez, Sandberg, Sandeen, Sanders,
  Sassolas, Sathyaprakash, Saulson, Sauter, Savage, Sawadsky, Schale, Scheuer,
  Schmidt, Schmidt, Schmidt, Schnabel, Schofield, Sch\"onbeck, Schreiber,
  Schuette, Schulte, Schutz, Schwalbe, Scott, Scott, Seidel, Sellers, Sengupta,
  Sentenac, Sequino, Sergeev, Shaddock, Shaffer, Shah, Shahriar, Shao, Shapiro,
  Shawhan, Sheperd, Shoemaker, Shoemaker, Siellez, Siemens, Sieniawska, Sigg,
  Silva, Singer, Singer, Singh, Singh, Singhal, Sintes, Slagmolen, Smith,
  Smith, Smith, Son, Sonnenberg, Sorazu, Sorrentino, Souradeep, Spencer,
  Srivastava, Staley, Steinke, Steinlechner, Steinlechner, Steinmeyer,
  Stephens, Stevenson, Stone, Strain, Stratta, Strigin, Sturani, Stuver,
  Summerscales, Sun, Sunil, Sutton, Swinkels, Szczepa\ifmmode~\acute{n}\else
  \'{n}\fi{}czyk, Tacca, Talukder, Tanner, T\'apai, Taracchini, Taylor, Taylor,
  Theeg, Thomas, Thomas, Thomas, Thorne, Thorne, Thrane, Tiwari, Tiwari,
  Tokmakov, Toland, Tonelli, Tornasi, Torrie, T\"oyr\"a, Travasso, Traylor,
  Trifir\`o, Trinastic, Tringali, Trozzo, Tsang, Tse, Tso, Tuyenbayev, Ueno,
  Ugolini, Unnikrishnan, Urban, Usman, Vahi, Vahlbruch, Vajente, Valdes,
  Vallisneri, van Bakel, van Beuzekom, van~den Brand, Van Den~Broeck,
  Vander-Hyde, van~der Schaaf, van Heijningen, van Veggel, Vardaro, Varma,
  Vass, Vas\'uth, Vecchio, Vedovato, Veitch, Veitch, Venkateswara, Venugopalan,
  Verkindt, Vetrano, Vicer\'e, Viets, Vinciguerra, Vine, Vinet, Vitale, Vo,
  Vocca, Vorvick, Voss, Vousden, Vyatchanin, Wade, Wade, Wade, Wald, Walet,
  Walker, Wallace, Walsh, Wang, Wang, Wang, Wang, Wang, Wang, Ward, Warner,
  Was, Watchi, Weaver, Wei, Weinert, Weinstein, Weiss, Wen, Wessel, We\ss{}els,
  Westphal, Wette, Whelan, Whiting, Whittle, Williams, Williams, Williamson,
  Willis, Willke, Wimmer, Winkler, Wipf, Wittel, Woan, Woehler, Wofford, Wong,
  Worden, Wright, Wu, Wu, Yam, Yamamoto, Yancey, Yap, Yu, Yu, Yvert,
  Zadro\ifmmode~\dot{z}\else \.{z}\fi{}ny, Zanolin, Zelenova, Zendri, Zevin,
  Zhang, Zhang, Zhang, Zhang, Zhao, Zhou, Zhou, Zhu, Zimmerman, Zucker, \&
  Zweizig}]{PhysRevLett.118.221101}
---. 2017, Phys. Rev. Lett., 118, 221101.
\newblock \url{https://link.aps.org/doi/10.1103/PhysRevLett.118.221101}

\bibitem[{{Alam} {et~al.}(2015){Alam}, {Albareti}, {Allende Prieto}, {Anders},
  {Anderson}, {Anderton}, {Andrews}, {Armengaud}, {Aubourg}, {Bailey}, \&
  et~al.}]{SDSS-DR12}
{Alam}, S., {Albareti}, F.~D., {Allende Prieto}, C., {et~al.} 2015, \apjs, 219,
  12

\bibitem[{{Baldassare} {et~al.}(2018){Baldassare}, {Geha}, \&
  {Greene}}]{2018ApJ...868..152B}
{Baldassare}, V.~F., {Geha}, M., \& {Greene}, J. 2018, \apj, 868, 152

\bibitem[{{Baldwin} {et~al.}(1981){Baldwin}, {Phillips}, \&
  {Terlevich}}]{1981PASP...93....5B}
{Baldwin}, J.~A., {Phillips}, M.~M., \& {Terlevich}, R. 1981, Publications of
  the Astronomical Society of the Pacific, 93, 5

\bibitem[{{Becker} {et~al.}(1994){Becker}, {White}, \&
  {Helfand}}]{1994ASPC...61..165B}
{Becker}, R.~H., {White}, R.~L., \& {Helfand}, D.~J. 1994, in Astronomical
  Society of the Pacific Conference Series, Vol.~61, Astronomical Data Analysis
  Software and Systems III, ed. D.~R. {Crabtree}, R.~J. {Hanisch}, \&
  J.~{Barnes}, 165

\bibitem[{{Cabrera-Vives} {et~al.}(2017){Cabrera-Vives}, {Reyes},
  {F{\"o}rster}, {Est{\'e}vez}, \& {Maureira}}]{2017ApJ...836...97C}
{Cabrera-Vives}, G., {Reyes}, I., {F{\"o}rster}, F., {Est{\'e}vez}, P.~A., \&
  {Maureira}, J.-C. 2017, \apj, 836, 97

\bibitem[{{Cann} {et~al.}(2019){Cann}, {Satyapal}, {Abel}, {Blecha},
  {Mushotzky}, {Reynolds}, \& {Secrest}}]{2019ApJ...870L...2C}
{Cann}, J.~M., {Satyapal}, S., {Abel}, N.~P., {et~al.} 2019, \apj, 870, L2

\bibitem[{{Chilingarian} {et~al.}(2018){Chilingarian}, {Katkov}, {Zolotukhin},
  {Grishin}, {Beletsky}, {Boutsia}, \& {Osip}}]{2018ApJ...863....1C}
{Chilingarian}, I.~V., {Katkov}, I.~Y., {Zolotukhin}, I.~Y., {et~al.} 2018,
  \apj, 863, 1

\bibitem[{{Cid Fernandes} {et~al.}(2005){Cid Fernandes}, {Mateus}, {Sodr{\'e}},
  {Stasi{\'n}ska}, \& {Gomes}}]{2005MNRAS.358..363C}
{Cid Fernandes}, R., {Mateus}, A., {Sodr{\'e}}, L., {Stasi{\'n}ska}, G., \&
  {Gomes}, J.~M. 2005, \mnras, 358, 363

\bibitem[{{Civano} {et~al.}(2012){Civano}, {Elvis}, {Brusa}, {Comastri},
  {Salvato}, {Zamorani}, {Aldcroft}, {Bongiorno}, {Capak}, {Cappelluti},
  {Cisternas}, {Fiore}, {Fruscione}, {Hao}, {Kartaltepe}, {Koekemoer}, {Gilli},
  {Impey}, {Lanzuisi}, {Lusso}, {Mainieri}, {Miyaji}, {Lilly}, {Masters},
  {Puccetti}, {Schawinski}, {Scoville}, {Silverman}, {Trump}, {Urry},
  {Vignali}, \& {Wright}}]{2012ApJS..201...30C}
{Civano}, F., {Elvis}, M., {Brusa}, M., {et~al.} 2012, The Astrophysical
  Journal Supplement Series, 201, 30

\bibitem[{{Dewangan} {et~al.}(2008){Dewangan}, {Mathur}, {Griffiths}, \&
  {Rao}}]{2008ApJ...689..762D}
{Dewangan}, G.~C., {Mathur}, S., {Griffiths}, R.~E., \& {Rao}, A.~R. 2008,
  \apj, 689, 762

\bibitem[{{Dong} {et~al.}(2012){Dong}, {Ho}, {Yuan}, {Wang}, {Fan}, {Zhou}, \&
  {Jiang}}]{2012ApJ...755..167D}
{Dong}, X.-B., {Ho}, L.~C., {Yuan}, W., {et~al.} 2012, \apj, 755, 167

\bibitem[{{Edri} {et~al.}(2012){Edri}, {Rafter}, {Chelouche}, {Kaspi}, \&
  {Behar}}]{2012ApJ...756...73E}
{Edri}, H., {Rafter}, S.~E., {Chelouche}, D., {Kaspi}, S., \& {Behar}, E. 2012,
  \apj, 756, 73

\bibitem[{{Eisenstein} {et~al.}(2001){Eisenstein}, {Annis}, {Gunn}, {Szalay},
  {Connolly}, {Nichol}, {Bahcall}, {Bernardi}, {Burles}, {Castander},
  {Fukugita}, {Hogg}, {Ivezi{\'c}}, {Knapp}, {Lupton}, {Narayanan}, {Postman},
  {Reichart}, {Richmond}, {Schneider}, {Schlegel}, {Strauss}, {SubbaRao},
  {Tucker}, {Vanden Berk}, {Vogeley}, {Weinberg}, \&
  {Yanny}}]{2001AJ....122.2267E}
{Eisenstein}, D.~J., {Annis}, J., {Gunn}, J.~E., {et~al.} 2001, \aj, 122, 2267

\bibitem[{{Falcke} {et~al.}(2004){Falcke}, {K{\"o}rding}, \&
  {Markoff}}]{2004A&A...414..895F}
{Falcke}, H., {K{\"o}rding}, E., \& {Markoff}, S. 2004, \aap, 414, 895

\bibitem[{{Filippenko} \& {Ho}(2003)}]{2003ApJ...588L..13F}
{Filippenko}, A.~V., \& {Ho}, L.~C. 2003, \apjl, 588, L13

\bibitem[{{F{\"o}rster} {et~al.}(2016){F{\"o}rster}, {Maureira}, {San
  Mart{\'\i}n}, {Hamuy}, {Mart{\'\i}nez}, {Huijse}, {Cabrera}, {Galbany}, {de
  Jaeger}, {Gonz{\'a}lez─Gait{\'a}n}, {Anderson}, {Kunkarayakti}, {Pignata},
  {Bufano}, {Litt{\'\i}n}, {Olivares}, {Medina}, {Smith}, {Vivas},
  {Est{\'e}vez}, {Mu{\~n}oz}, \& {Vera}}]{2016ApJ...832..155F}
{F{\"o}rster}, F., {Maureira}, J.~C., {San Mart{\'\i}n}, J., {et~al.} 2016,
  \apj, 832, 155

\bibitem[{{F{\"o}rster} {et~al.}(2018){F{\"o}rster}, {Moriya}, {Maureira},
  {Anderson}, {Blinnikov}, {Bufano}, {Cabrera-Vives}, {Clocchiatti}, {de
  Jaeger}, {Est{\'e}vez}, {Galbany}, {Gonz{\'a}lez- Gait{\'a}n},
  {Gr{\"a}fener}, {Hamuy}, {Hsiao}, {Huentelemu}, {Huijse}, {Kuncarayakti},
  {Mart{\'\i}nez}, {Medina}, {Olivares E.}, {Pignata}, {Razza}, {Reyes}, {San
  Mart{\'\i}n}, {Smith}, {Vera}, {Vivas}, {de Ugarte Postigo}, {Yoon},
  {Ashall}, {Fraser}, {Gal-Yam}, {Kankare}, {Le Guillou}, {Mazzali}, {Walton},
  \& {Young}}]{2018NatAs.tmp..122F}
{F{\"o}rster}, F., {Moriya}, T.~J., {Maureira}, J.~C., {et~al.} 2018, Nature
  Astronomy, 122

\bibitem[{{Greene} \& {Ho}(2004)}]{2004ApJ...610..722G}
{Greene}, J.~E., \& {Ho}, L.~C. 2004, \apj, 610, 722

\bibitem[{{Greene} \& {Ho}(2007)}]{2007ApJ...670...92G}
---. 2007, \apj, 670, 92

\bibitem[{{Kamizasa} {et~al.}(2012){Kamizasa}, {Terashima}, \&
  {Awaki}}]{2012ApJ...751...39K}
{Kamizasa}, N., {Terashima}, Y., \& {Awaki}, H. 2012, \apj, 751, 39

\bibitem[{{Kewley} {et~al.}(2006){Kewley}, {Groves}, {Kauffmann}, \&
  {Heckman}}]{2006MNRAS.372..961K}
{Kewley}, L.~J., {Groves}, B., {Kauffmann}, G., \& {Heckman}, T. 2006, \mnras,
  372, 961

\bibitem[{{Kormendy} \& {Ho}(2013)}]{2013ARA&A..51..511K}
{Kormendy}, J., \& {Ho}, L.~C. 2013, Annual Review of Astronomy and
  Astrophysics, 51, 511

\bibitem[{{Li} {et~al.}(2007){Li}, {Hernquist}, {Robertson}, {Cox}, {Hopkins},
  {Springel}, {Gao}, {Di Matteo}, {Zentner}, {Jenkins}, \&
  {Yoshida}}]{2007ApJ...665..187L}
{Li}, Y., {Hernquist}, L., {Robertson}, B., {et~al.} 2007, \apj, 665, 187

\bibitem[{{Liu} {et~al.}(2018){Liu}, {Yuan}, {Dong}, {Zhou}, \&
  {Liu}}]{2018ApJS..235...40L}
{Liu}, H.-Y., {Yuan}, W., {Dong}, X.-B., {Zhou}, H., \& {Liu}, W.-J. 2018, The
  Astrophysical Journal Supplement Series, 235, 40

\bibitem[{{Maraston} {et~al.}(2009){Maraston}, {Str{\"o}mb{\"a}ck}, {Thomas},
  {Wake}, \& {Nichol}}]{2009MNRAS.394L.107M}
{Maraston}, C., {Str{\"o}mb{\"a}ck}, G., {Thomas}, D., {Wake}, D.~A., \&
  {Nichol}, R.~C. 2009, \mnras, 394, L107

\bibitem[{{Mart{\'\i}nez-Palomera} {et~al.}(2018){Mart{\'\i}nez-Palomera},
  {F{\"o}rster}, {Protopapas}, {Maureira}, {Lira}, {Cabrera-Vives}, {Huijse},
  {Galbany}, {de Jaeger}, {Gonz{\'a}lez- Gait{\'a}n}, {Medina}, {Pignata}, {San
  Mart{\'\i}n}, {Hamuy}, \& {Mu{\~n}oz}}]{2018AJ....156..186M}
{Mart{\'\i}nez-Palomera}, J., {F{\"o}rster}, F., {Protopapas}, P., {et~al.}
  2018, \aj, 156, 186

\bibitem[{{Matthews} \& {Sandage}(1963)}]{1963ApJ...138...30M}
{Matthews}, T.~A., \& {Sandage}, A.~R. 1963, \apj, 138, 30

\bibitem[{{Medina} {et~al.}(2017){Medina}, {Mu{\~n}oz}, {Vivas}, {F{\"o}rster},
  {Carlin}, {Martinez}, {Galbany}, {Gonz{\'a}lez-Gait{\'a}n}, {Hamuy}, {de
  Jaeger}, {Maureira}, \& {San Mart{\'\i}n}}]{2017ApJ...845L..10M}
{Medina}, G.~E., {Mu{\~n}oz}, R.~R., {Vivas}, A.~K., {et~al.} 2017, \apj, 845,
  L10

\bibitem[{{Medina} {et~al.}(2018){Medina}, {Mu{\~n}oz}, {Vivas}, {Carlin},
  {F{\"o}rster}, {Mart{\'\i}nez}, {Galbany}, {Gonz{\'a}lez-Gait{\'a}n},
  {Hamuy}, {de Jaeger}, {Maureira}, \& {San Mart{\'\i}n}}]{2018ApJ...855...43M}
---. 2018, \apj, 855, 43

\bibitem[{{Merloni} {et~al.}(2003){Merloni}, {Heinz}, \& {di
  Matteo}}]{2003MNRAS.345.1057M}
{Merloni}, A., {Heinz}, S., \& {di Matteo}, T. 2003, \mnras, 345, 1057

\bibitem[{{Mezcua}(2017)}]{2017IJMPD..2630021M}
{Mezcua}, M. 2017, International Journal of Modern Physics D, 26, 1730021

\bibitem[{{Miyoshi} {et~al.}(1995){Miyoshi}, {Moran}, {Herrnstein},
  {Greenhill}, {Nakai}, {Diamond}, \& {Inoue}}]{1995Natur.373..127M}
{Miyoshi}, M., {Moran}, J., {Herrnstein}, J., {et~al.} 1995, \nat, 373, 127

\bibitem[{{Mortlock} {et~al.}(2011){Mortlock}, {Warren}, {Venemans}, {Patel},
  {Hewett}, {McMahon}, {Simpson}, {Theuns}, {Gonz{\'a}les- Solares}, {Adamson},
  {Dye}, {Hambly}, {Hirst}, {Irwin}, {Kuiper}, {Lawrence}, \&
  {R{\"o}ttgering}}]{2011Natur.474..616M}
{Mortlock}, D.~J., {Warren}, S.~J., {Venemans}, B.~P., {et~al.} 2011, \nat,
  474, 616

\bibitem[{{Pe{\~n}a} {et~al.}(2018){Pe{\~n}a}, {Fuentes}, {F{\"o}rster},
  {Maureira}, {San Mart{\'\i}n}, {Litt{\'\i}n}, {Huijse}, {Cabrera-Vives},
  {Est{\'e}vez}, {Galbany}, {Gonz{\'a}lez-Gait{\'a}n}, {Mart{\'\i}nez}, {de
  Jaeger}, \& {Hamuy}}]{2018AJ....155..135P}
{Pe{\~n}a}, J., {Fuentes}, C., {F{\"o}rster}, F., {et~al.} 2018, \aj, 155, 135

\bibitem[{{Peterson}(2001)}]{2001sac..conf....3P}
{Peterson}, B.~M. 2001, in Advanced Lectures on the Starburst-AGN, 3

\bibitem[{{Plotkin} {et~al.}(2012){Plotkin}, {Markoff}, {Kelly}, {K{\"o}rding},
  \& {Anderson}}]{2012MNRAS.419..267P}
{Plotkin}, R.~M., {Markoff}, S., {Kelly}, B.~C., {K{\"o}rding}, E., \&
  {Anderson}, S.~F. 2012, \mnras, 419, 267

\bibitem[{{Reines} \& {Comastri}(2016)}]{2016PASA...33...54R}
{Reines}, A.~E., \& {Comastri}, A. 2016, Publications of the Astronomical
  Society of Australia, 33, e054

\bibitem[{{Reines} {et~al.}(2013){Reines}, {Greene}, \&
  {Geha}}]{2013ApJ...775..116R}
{Reines}, A.~E., {Greene}, J.~E., \& {Geha}, M. 2013, \apj, 775, 116

\bibitem[{{Remillard} \& {McClintock}(2006)}]{2006ARA&A..44...49R}
{Remillard}, R.~A., \& {McClintock}, J.~E. 2006, Annual Review of Astronomy and
  Astrophysics, 44, 49

\bibitem[{{Rosen} {et~al.}(2016){Rosen}, {Webb}, {Watson}, {Ballet}, {Barret},
  {Braito}, {Carrera}, {Ceballos}, {Coriat}, {Della Ceca}, {Denkinson},
  {Esquej}, {Farrell}, {Freyberg}, {Gris{\'e}}, {Guillout}, {Heil},
  {Koliopanos}, {Law-Green}, {Lamer}, {Lin}, {Martino}, {Michel}, {Motch},
  {Nebot Gomez-Moran}, {Page}, {Page}, {Page}, {Pakull}, {Pye}, {Read},
  {Rodriguez}, {Sakano}, {Saxton}, {Schwope}, {Scott}, {Sturm}, {Traulsen},
  {Yershov}, \& {Zolotukhin}}]{2016A&A...590A...1R}
{Rosen}, S.~R., {Webb}, N.~A., {Watson}, M.~G., {et~al.} 2016, \aap, 590, A1

\bibitem[{{S{\'a}nchez} {et~al.}(2017){S{\'a}nchez}, {Lira}, {Cartier},
  {P{\'e}rez}, {Miranda}, {Yovaniniz}, {Ar{\'e}valo}, {Milvang-Jensen},
  {Fynbo}, {Dunlop}, {Coppi}, \& {Marchesi}}]{2017ApJ...849..110S}
{S{\'a}nchez}, P., {Lira}, P., {Cartier}, R., {et~al.} 2017, \apj, 849, 110

\bibitem[{{Sch{\"o}del} {et~al.}(2002){Sch{\"o}del}, {Ott}, {Genzel},
  {Hofmann}, {Lehnert}, {Eckart}, {Mouawad}, {Alexander}, {Reid}, {Lenzen},
  {Hartung}, {Lacombe}, {Rouan}, {Gendron}, {Rousset}, {Lagrange}, {Brandner},
  {Ageorges}, {Lidman}, {Moorwood}, {Spyromilio}, {Hubin}, \&
  {Menten}}]{2002Natur.419..694S}
{Sch{\"o}del}, R., {Ott}, T., {Genzel}, R., {et~al.} 2002, \nat, 419, 694

\bibitem[{{Shakura} \& {Sunyaev}(1973)}]{1973A&A....24..337S}
{Shakura}, N.~I., \& {Sunyaev}, R.~A. 1973, \aap, 24, 337

\bibitem[{{Strauss} {et~al.}(2002){Strauss}, {Weinberg}, {Lupton}, {Narayanan},
  {Annis}, {Bernardi}, {Blanton}, {Burles}, {Connolly}, {Dalcanton}, {Doi},
  {Eisenstein}, {Frieman}, {Fukugita}, {Gunn}, {Ivezi{\'c}}, {Kent}, {Kim},
  {Knapp}, {Kron}, {Munn}, {Newberg}, {Nichol}, {Okamura}, {Quinn}, {Richmond},
  {Schlegel}, {Shimasaku}, {SubbaRao}, {Szalay}, {Vanden Berk}, {Vogeley},
  {Yanny}, {Yasuda}, {York}, \& {Zehavi}}]{2002AJ....124.1810S}
{Strauss}, M.~A., {Weinberg}, D.~H., {Lupton}, R.~H., {et~al.} 2002, \aj, 124,
  1810

\bibitem[{{Strubbe} \& {Quataert}(2009)}]{2009MNRAS.400.2070S}
{Strubbe}, L.~E., \& {Quataert}, E. 2009, \mnras, 400, 2070

\bibitem[{{Trump} {et~al.}(2015){Trump}, {Sun}, {Zeimann}, {Luck}, {Bridge},
  {Grier}, {Hagen}, {Juneau}, {Montero-Dorta}, {Rosario}, {Brandt},
  {Ciardullo}, \& {Schneider}}]{2015ApJ...811...26T}
{Trump}, J.~R., {Sun}, M., {Zeimann}, G.~R., {et~al.} 2015, \apj, 811, 26

\bibitem[{{Valdes} {et~al.}(2014){Valdes}, {Gruendl}, \& {DES
  Project}}]{2014ASPC..485..379V}
{Valdes}, F., {Gruendl}, R., \& {DES Project}. 2014, in Astronomical Data
  Analysis Software and Systems XXIII, Vol. 485, 379

\bibitem[{{Volonteri}(2010)}]{2010A&ARv..18..279V}
{Volonteri}, M. 2010, Astronomy and Astrophysics Review, 18, 279

\bibitem[{{Willott} {et~al.}(2007){Willott}, {Delorme}, {Omont}, {Bergeron},
  {Delfosse}, {Forveille}, {Albert}, {Reyl{\'e}}, {Hill}, {Gully-Santiago},
  {Vinten}, {Crampton}, {Hutchings}, {Schade}, {Simard}, {Sawicki}, {Beelen},
  \& {Cox}}]{2007AJ....134.2435W}
{Willott}, C.~J., {Delorme}, P., {Omont}, A., {et~al.} 2007, \aj, 134, 2435

\bibitem[{{Yoo} {et~al.}(2007){Yoo}, {Miralda-Escud{\'e}}, {Weinberg}, {Zheng},
  \& {Morgan}}]{2007ApJ...667..813Y}
{Yoo}, J., {Miralda-Escud{\'e}}, J., {Weinberg}, D.~H., {Zheng}, Z., \&
  {Morgan}, C.~W. 2007, \apj, 667, 813

\bibitem[{{Yu} \& {Tremaine}(2002)}]{2002MNRAS.335..965Y}
{Yu}, Q., \& {Tremaine}, S. 2002, \mnras, 335, 965

\end{thebibliography}

\appendix
\section{Examples of Light Curves and Image Sequences} \label{apx: img_lc_ex}

The following figures present image sequences and \textit{g} band light curves of a sample of variable galaxies.
Top panels show the image sequences where the first row are original unconvolved images, in the second row are convolved images as well as the adopted photometric aperture (red outer circle), and in the third row are the convolution kernels used for PSF matching. Yellow numbers are MJD of each stamp. Note that for reference epoch, the first two rows match while the last row shows the modeled PSF.
Bottom panels show the light curve of the source. Black points are the observed magnitudes with corresponding uncertainties. Red dashed lines represent the 2 and 3 sigma intervals for non-variable galaxies in the field, while the gray shaded regions shows the 2 and 3 times the photometric uncertainties around the median magnitude.
\begin{figure*}[h!]
\gridline{\fig{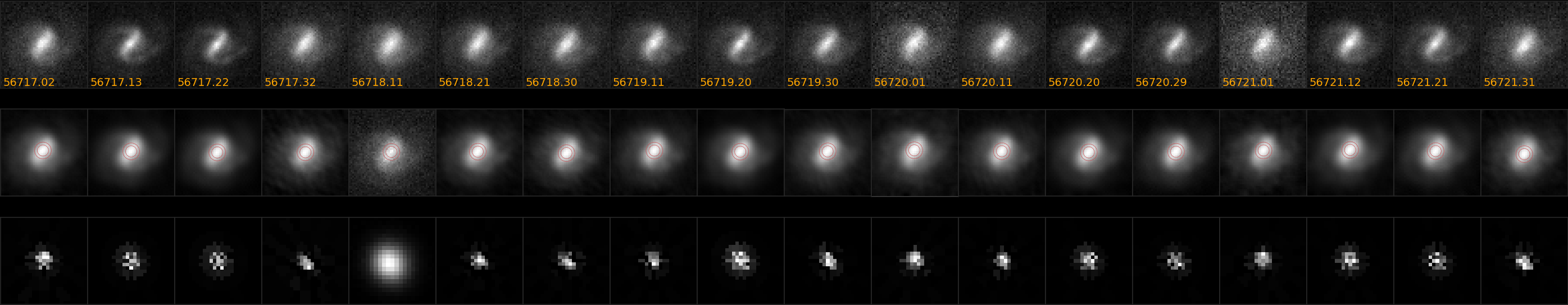}{0.8\textwidth}{}}
\vspace{-25px}
\gridline{\fig{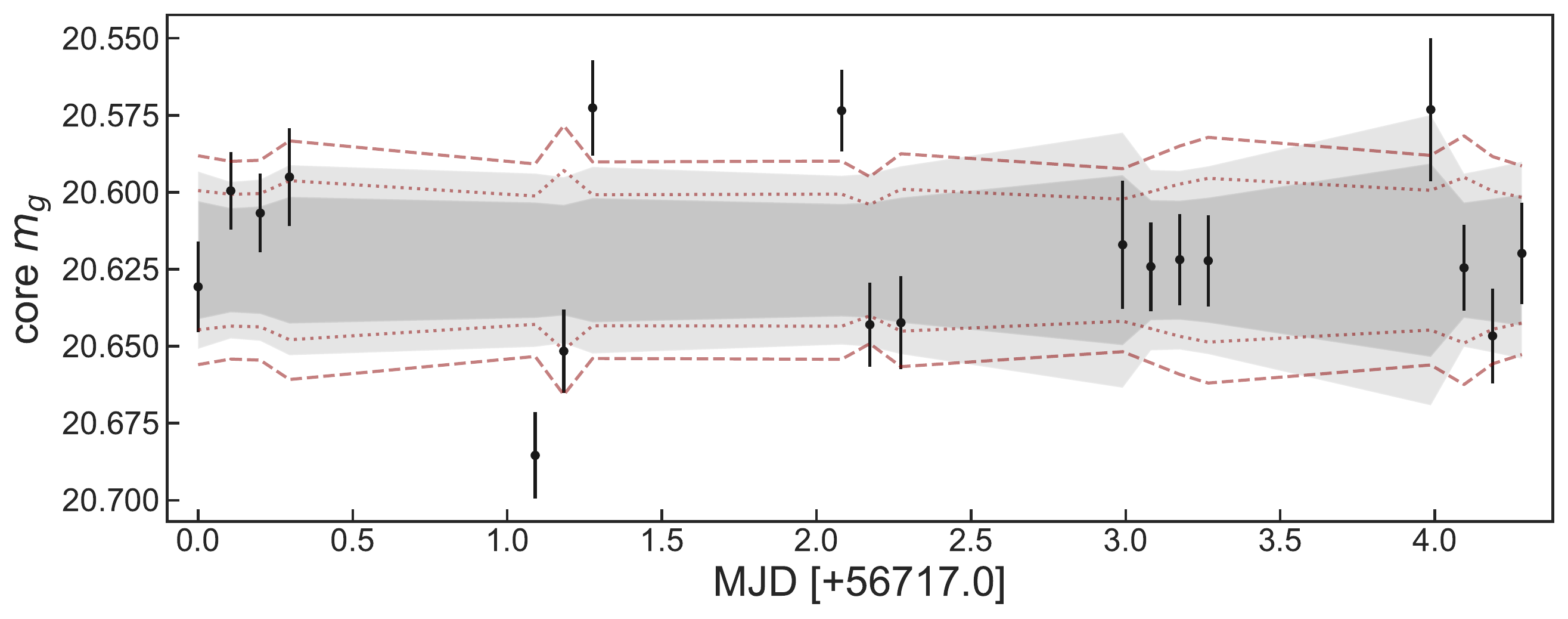}{0.6\textwidth}{}}
\vspace{-20px}
\caption{Spiral Galaxy, SDSS $m_g = 18.49$}
\end{figure*}
\begin{figure*}[h!]
\gridline{\fig{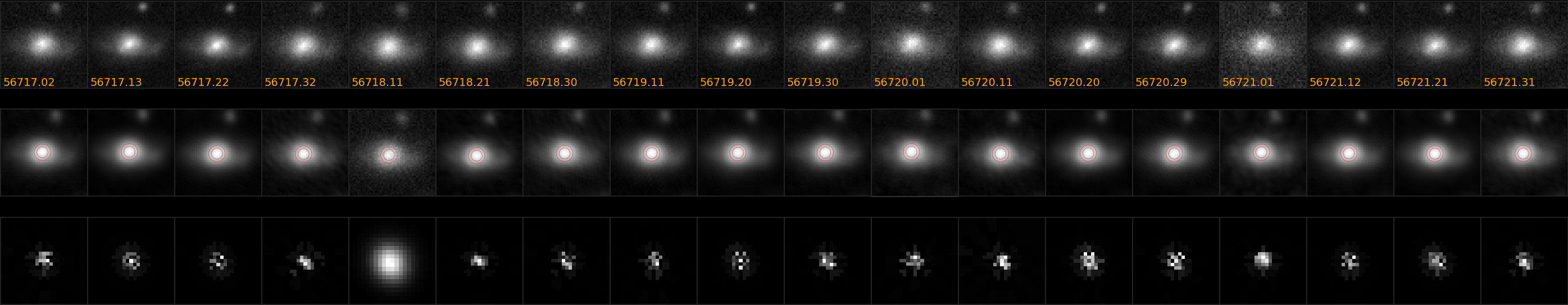}{0.8\textwidth}{}}
\vspace{-25px}
\gridline{\fig{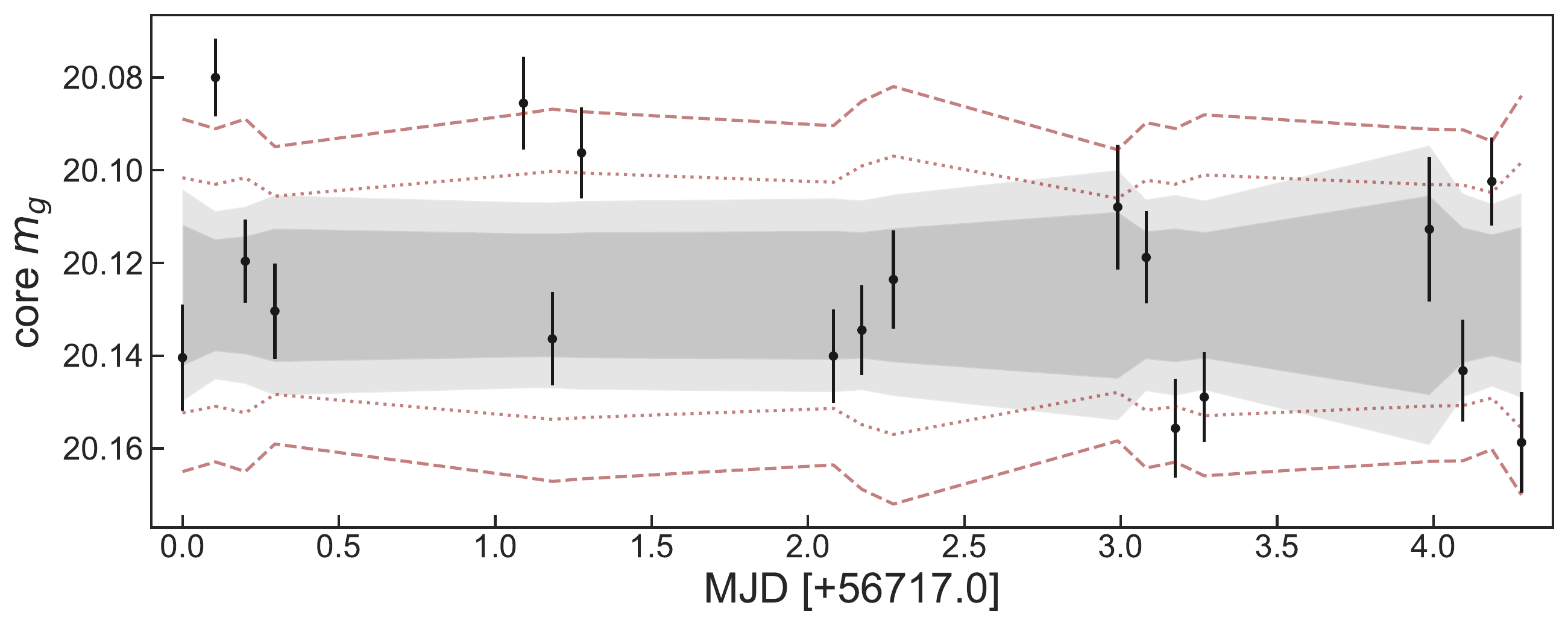}{0.6\textwidth}{}}
\vspace{-20px}
\caption{Spiral Galaxy, SDSS $m_g = 18.33$}
\end{figure*}
\begin{figure*}[h!]
\gridline{\fig{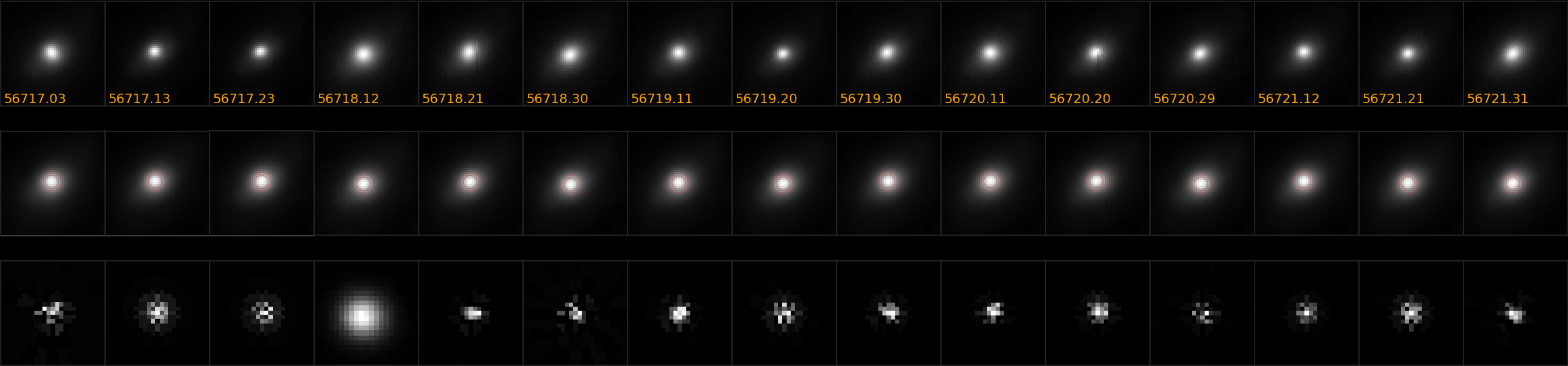}{0.8\textwidth}{}}
\vspace{-25px}
\gridline{\fig{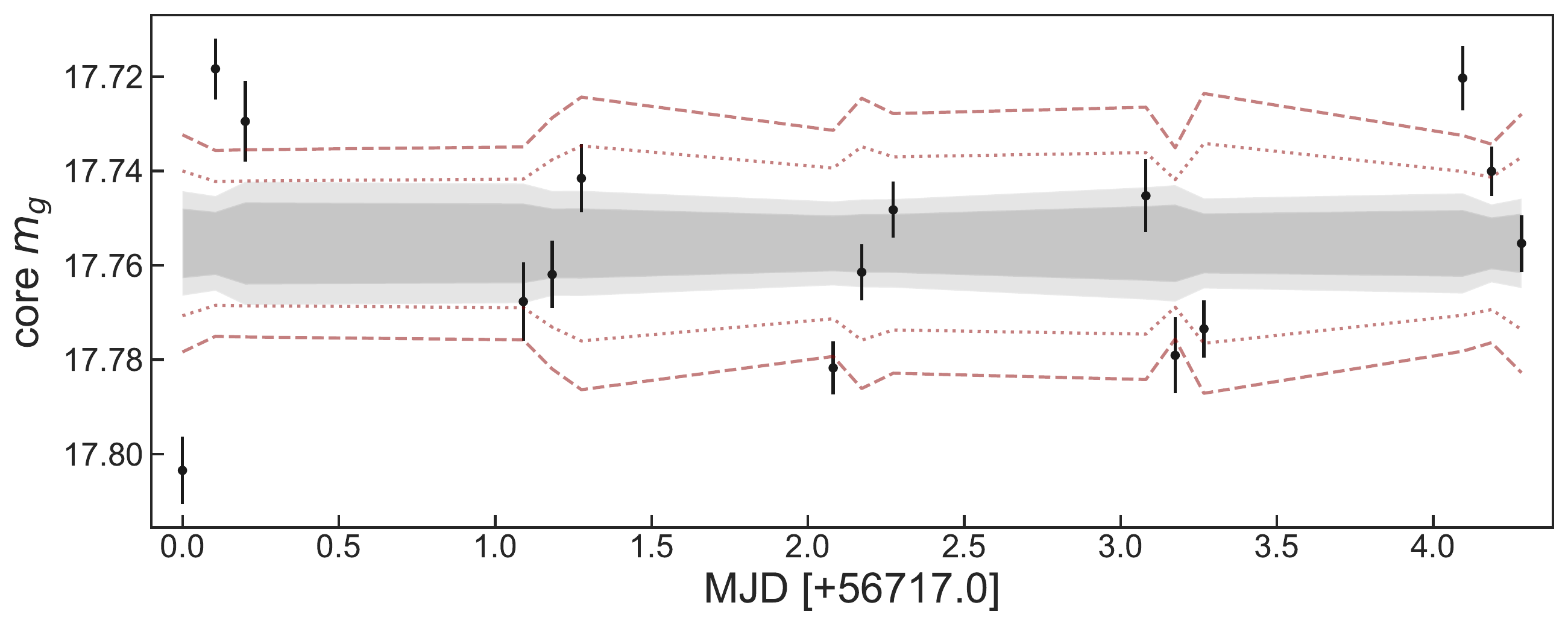}{0.6\textwidth}{}}
\vspace{-20px}
\caption{Elliptical Galaxy, SDSS $m_g = 15.62$}
\end{figure*}
\begin{figure*}[h!]
\gridline{\fig{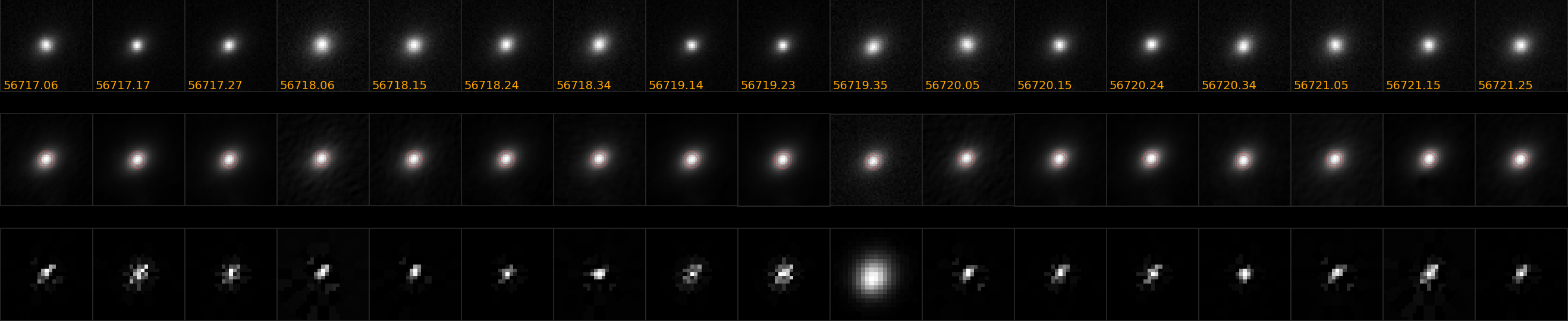}{0.8\textwidth}{}}
\vspace{-25px}
\gridline{\fig{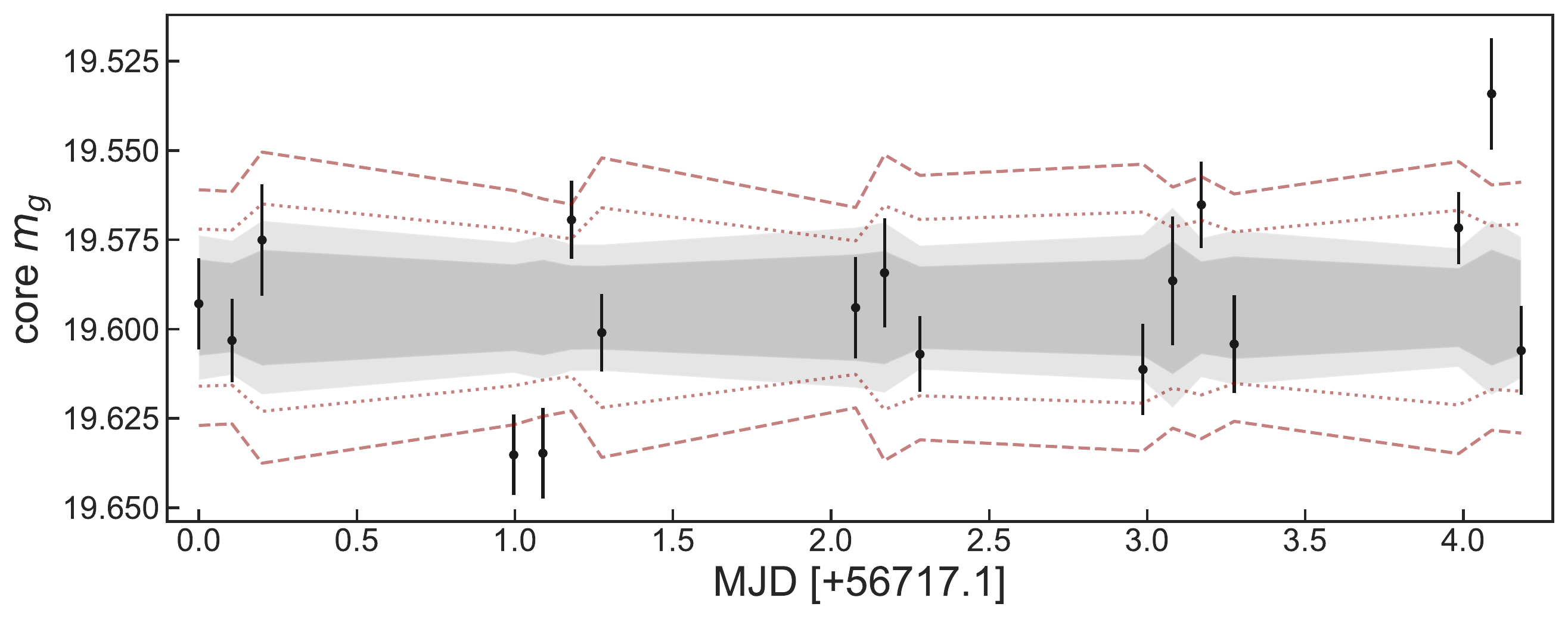}{0.6\textwidth}{}}
\vspace{-20px}
\caption{Elliptical Galaxy, SDSS $m_g = 18.26$}
\end{figure*}
\begin{figure*}[h!]
\gridline{\fig{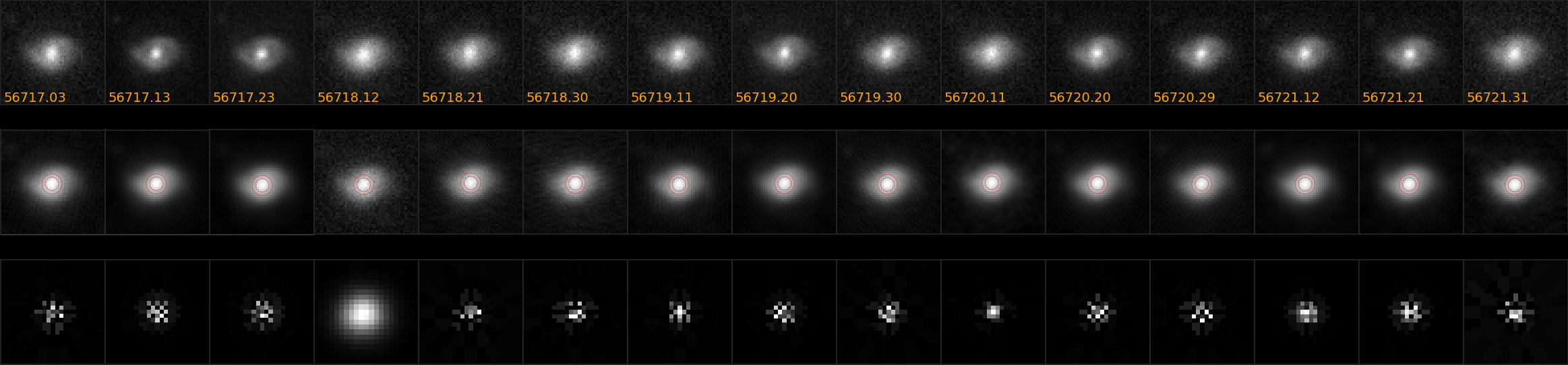}{0.8\textwidth}{}}
\vspace{-25px}
\gridline{\fig{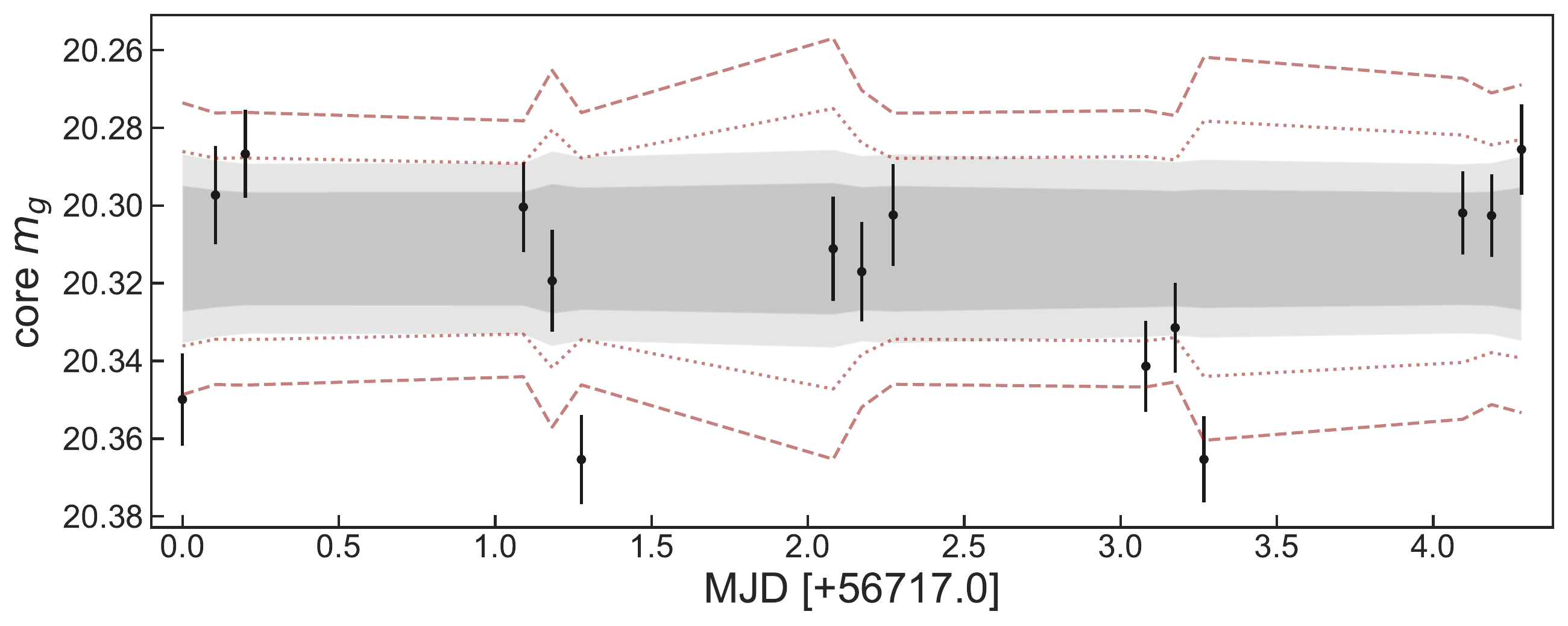}{0.6\textwidth}{}}
\vspace{-20px}
\caption{Spiral Galaxy, SDSS $m_g = 18.61$}
\end{figure*}
\begin{figure*}[h!]
\gridline{\fig{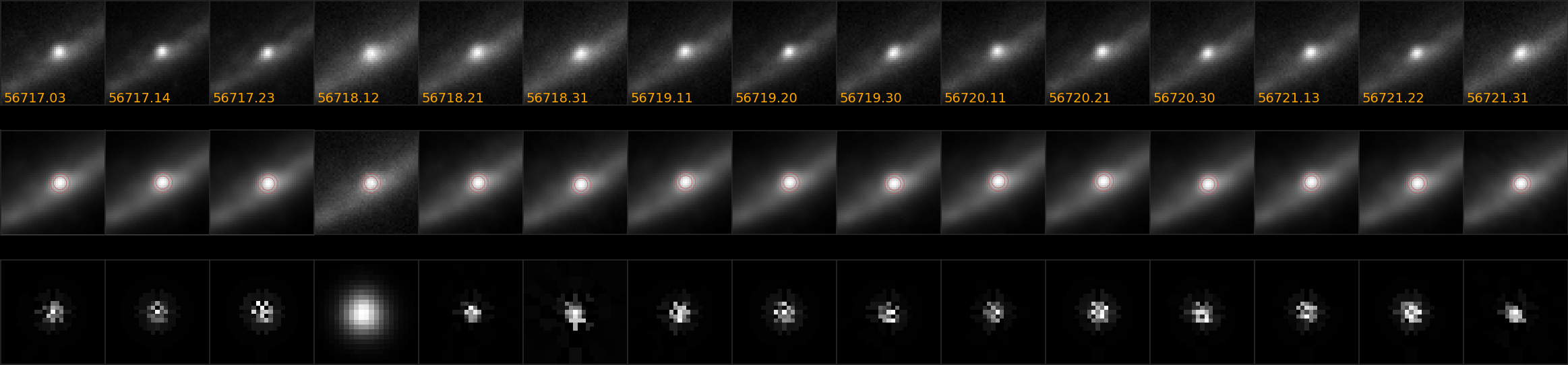}{0.8\textwidth}{}}
\vspace{-25px}
\gridline{\fig{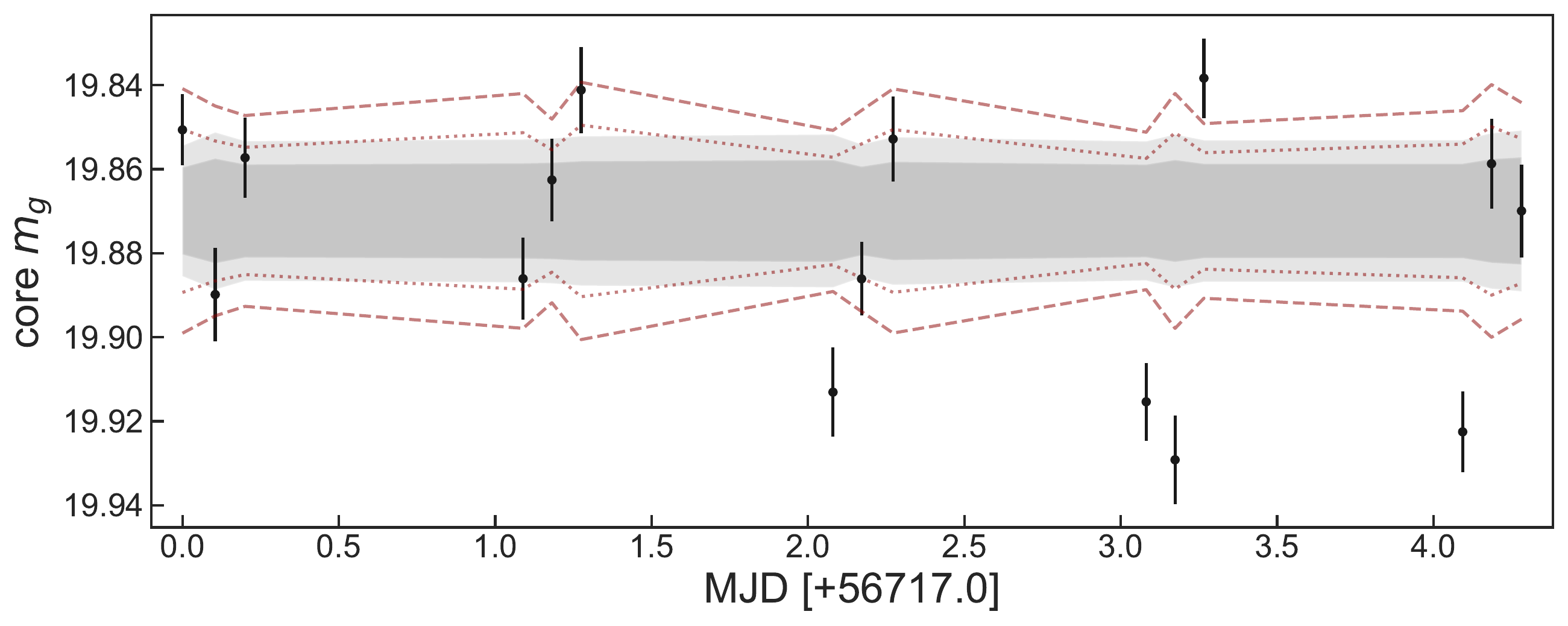}{0.6\textwidth}{}}
\vspace{-20px}
\caption{Spiral Galaxy, SDSS $m_g = 16.48$}
\end{figure*}
\begin{figure*}[h!]
\gridline{\fig{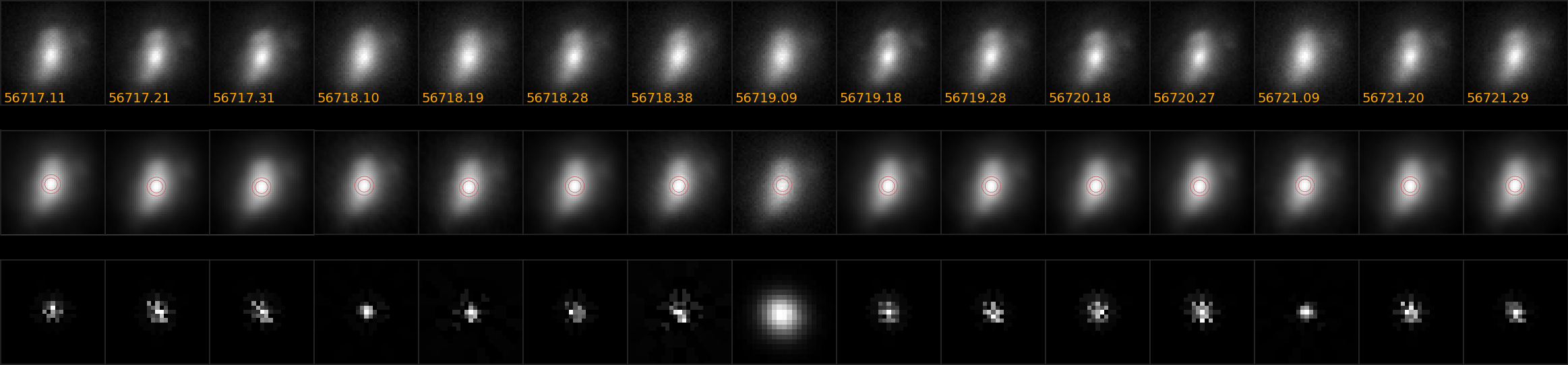}{0.8\textwidth}{}}
\vspace{-25px}
\gridline{\fig{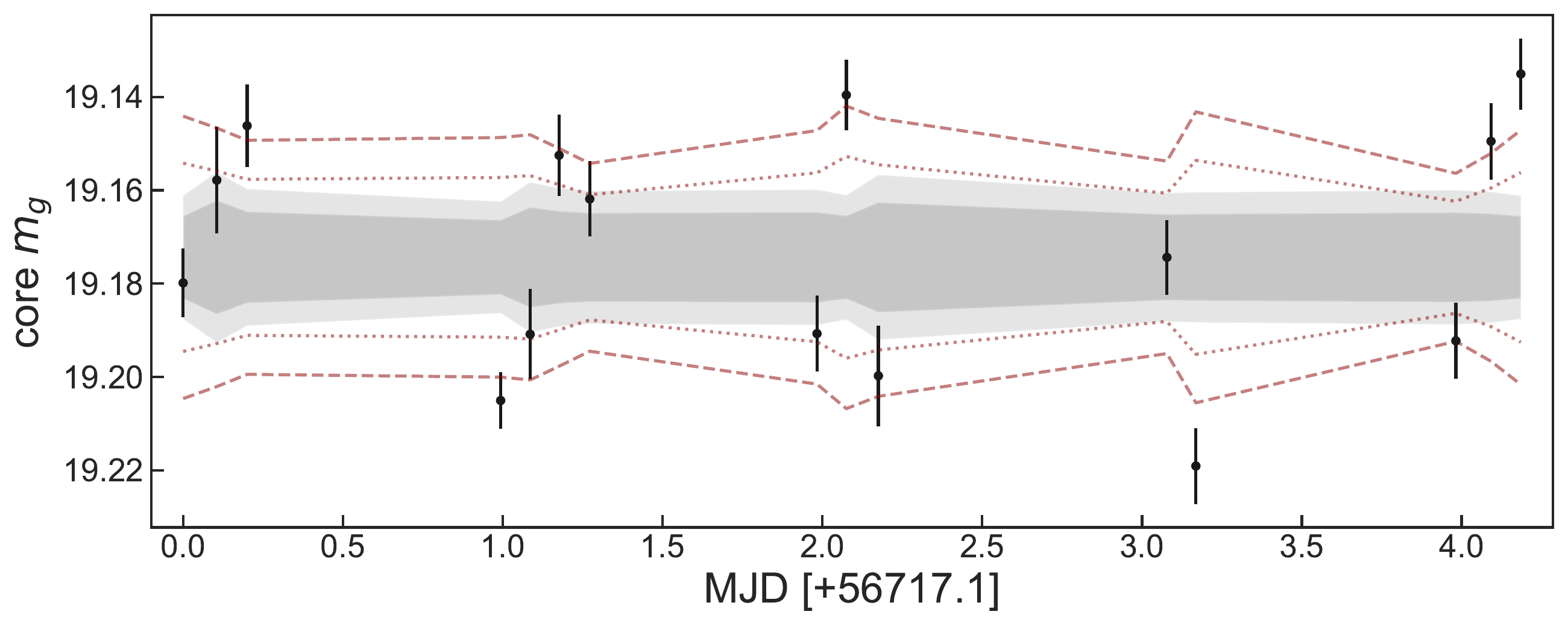}{0.6\textwidth}{}}
\vspace{-20px}
\caption{Irregular Galaxy, SDSS $m_g = 16.38$}
\end{figure*}
\begin{figure*}[h!]
\gridline{\fig{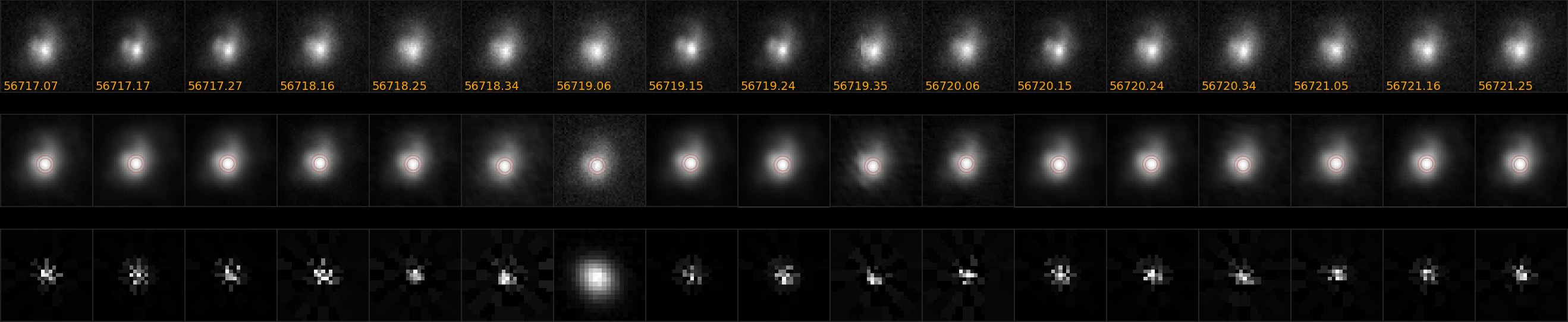}{0.8\textwidth}{}}
\vspace{-25px}
\gridline{\fig{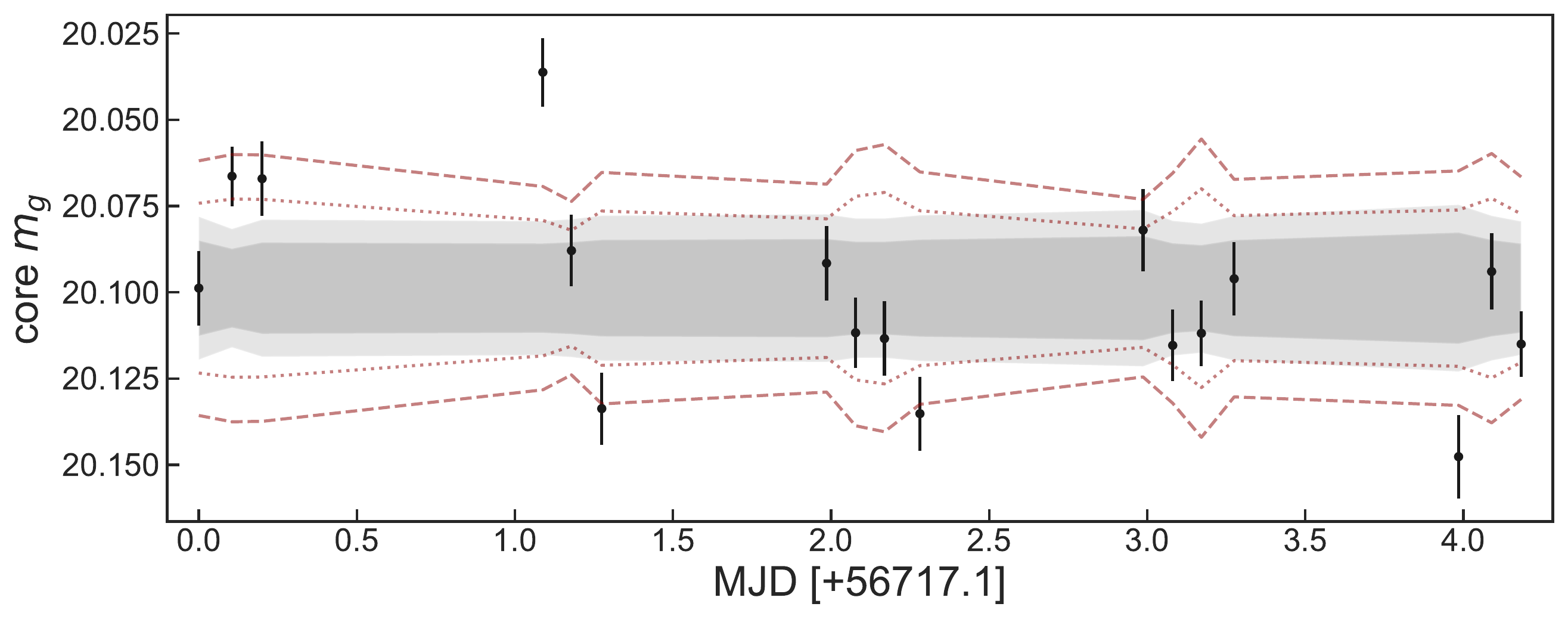}{0.6\textwidth}{}}
\vspace{-20px}
\caption{Spiral Galaxy, SDSS $m_g = 18.07$}
\end{figure*}

\end{document}